%% file: JLM_JDP_POF.tex
\documentclass[aip,preprint,pof]{revtex4-1}

\usepackage[english]{babel}
\usepackage{graphicx}
\usepackage{psfrag}
\usepackage{epsfig}
\usepackage{amssymb,amsmath,amsbsy}
\usepackage{color}
\usepackage{natbib}
\usepackage[percent]{overpic}

\begin{document}

\title{Role of Solution Conductivity in Reaction Induced Charge Auto-Electrophoresis}
\author{Jeffrey L. Moran\footnote{Present address: Department of Mechanical Engineering, Massachusetts Institute of Technology, Cambridge, MA 02139}}
\author{Jonathan D. Posner \footnote{Corresponding author.  Email: jposner$@$uw.edu}}

\affiliation{Department of Mechanical Engineering, University of Washington, Seattle, WA 98195-2600, USA}

\input{abstract.tex}

\maketitle

\section{Introduction}
\input{intro.tex}
\label{intro}
\section{Problem Formulation and Theory}
\input{theory.tex}
\label{theory}
\section{Results and Discussion}
\input{results.tex}
\label{results}
\section{Summary}
\input{summary.tex}
\label{summary}
\begin{acknowledgments}
This work was sponsored by NSF grant CBET-0853379. 
\end{acknowledgments}
\bibliography{nm}

\end{document}

%% file: abstract.tex

\begin{abstract}
Catalytic bimetallic Janus particles swim by a bipolar electrochemical propulsion mechanism that results from  electroosmotic fluid slip around the particle surface.  The flow is driven by electrical body forces which are generated from a coupling of a reaction-induced electric field and net charge in the diffuse layer surrounding the particle. This paper presents simulations, scaling, and physical descriptions of the experimentally observed trend that the swimming speed decays rapidly with increasing solution conductivity.  The simulations solve the full Poisson-Nernst-Planck-Stokes equations with multiple ionic species, a cylindrical particle in an infinite fluid, and nonlinear Butler-Volmer boundary conditions to represent the electrochemical surface reactions. The speed of bimetallic particles is reduced in high-conductivity solutions because of reductions in the induced electric field in the diffuse layer near the rod, the total reaction rate, and the magnitude of the rod zeta potential.  The results in this work suggest that the auto-electrophoretic mechanism is inherently susceptible to speed reductions in higher ionic strength solutions.
\end{abstract}

%% file: intro.tex
Bimetallic particles have attracted significant interest over the past decade due to their ability to move autonomously through aqueous solutions containing hydrogen peroxide (H$_2$O$_2$).\cite{paxton_catalytic_2004,wang_can_2009,ebbens_pursuit_2010,takagi_dispersion_2013}  Several proof-of-concept experiments have demonstrated that these nanomotors can perform potentially useful tasks at the micron scale, such as motion at over 100 body lengths per second,\cite{laocharoensuk_carbon-nanotube-induced_2008} cargo transport and delivery,\cite{burdick_synthetic_2008,sundararajan_catalytic_2008} electrochemically\cite{calvo-marzal_electrochemically-triggered_2009} and thermally controlled movement,\cite{balasubramanian_thermal_2009} and chemical sensing,\cite{kagan_chemical_2009} with possible future applications in targeted drug delivery, assembly of complex micro-scale structures, and the sensing and removal of chemical impurities in drinking water.  The potential utility of the particles for these applications may be aided by a more rigorous understanding of the physical principles governing their motion.

The most common realization of a swimming bimetallic particle is a solid rod, 2 microns in length and between 200-400 nanometers in diameter, consisting of two metallic segments.  The typical metals used are platinum (Pt) and gold (Au), although other metals have also been demonstrated.\cite{wang_bipolar_2006}  The self-propelled motion of bimetallic particles is notable in that it always occurs with the same end directed forward (Pt in the case of Pt/Au particles). We previously demonstrated that bimetallic spherical motors may also be fabricated by coating polystyrene spheres completely with Au, and then half-coating the spheres with Pt, forming Pt/Au Janus spheres.\cite{wheat_rapid_2010}

Several theories have been proposed to explain the operative mechanism of conversion of chemical to kinetic energy for bimetallic self-propelling particles, including gradients in interfacial tension,\cite{paxton_catalytic_2004} viscous Brownian ratchet,\cite{dhar_autonomously_2006} and bubble propulsion.\cite{nina_toward_2008}  In 2005, \citet{paxton_motility_2005} proposed a bipolar electrochemical propulsion mechanism that considers a Pt/Au rod as a short-circuited galvanic cell, with Pt acting as the anode and Au as the cathode.  In this mechanism, catalytic reduction and oxidation reactions occur on the Au and Pt surfaces, respectively, driving an electron current within the rod from Pt to Au and a corresponding proton current through the fluid surrounding the rod, also from Pt to Au.  The proton current is accompanied by an electric field pointing along the rod's axis, from Pt to Au.  The electric field drives fluid from Pt to Au and propels the negatively charged particle with the Pt end forward.

This mechanism is often referred to as \textit{self-electrophoresis} and was first considered in 1956 by Nobel laureate Peter Mitchell, who proposed that some species of microorganisms might move by generating ion currents in their bodies and the surrounding solution, establishing an electric field that propels the charged organism.\cite{mitchell_hypothetical_1956,mitchell_self-electrophoretic_1972}  Since Mitchell's original theory, we and others have provided more detailed scaling analyses, models, and simulations of electrophoretically powered swimmers.\cite{lammert_ion_1996,paxton_motility_2005,paxton_catalytically_2006,wang_bipolar_2006,moran_locomotion_2010,moran_electrokinetic_2011,sabass_nonlinear_2012}  For a more thorough review of previous efforts to physically understand the mechanism of electrophoretic self-propulsion, the interested reader is referred to our previous work.\cite{moran_electrokinetic_2011}

In this work, we present simulations, scaling, and physical descriptions of the experimentally observed trend that the swimming speed of bimetallic particles decays rapidly upon the addition of electrolyte to the hydrogen peroxide solution.  Many of the envisioned applications for the bimetallic particles would require them to swim at non-trivial speeds in conductive environments (e.g., biological media).  This salt-dependent reduction in swimming velocity was first shown in 2006 by Paxton \textit{et al.},\cite{paxton_catalytically_2006} who measured the speed of Pt/Au rods in 3.7 wt. \% hydrogen peroxide as a function of concentration of sodium nitrate (NaNO$_3$) and lithium nitrate (LiNO$_3$).  Their results suggested a linear correlation between swimming speed and solution resistivity (reciprocal of solution conductivity).  Similar to the result of \citet{lammert_ion_1996} for a spherical self-electrophoretic cell, Paxton \textit{et al.} proposed that the rod's swimming speed is characterized by the Helmholtz-Smoluchowski equation, $U = \mu_e E_0$, where $\mu_e$ is the rod's electrophoretic mobility and $E_0$ is a characteristic magnitude of the self-generated electric field.  Paxton \textit{et al.} then applied Ohm's Law and assumed that the electric field is given as $E_0 = i / \sigma$, where $i$ is the current density due to the electrochemical reactions, and $\sigma$ is the electrical conductivity of the solution.  Combining the Helmholtz-Smoluchowski and Ohm's Law equations, Paxton \textit{et al.} obtained an expression for the swimming velocity $U$ that varies inversely with solution conductivity, given as\cite{paxton_catalytically_2006}
\begin{equation}
U = \frac{\mu_e i}{\sigma}.
\label{HSPax}
\end{equation}

\citet{golestanian_designing_2007} derived formulae for the steady-state speed of several phoretic swimmers, including a Janus cylinder similar to the Pt/Au rods.  Their analytical result was similar in form to the Helmholtz-Smoluchowski-like expression above (\ref{HSPax}), but included correction terms to account for the non-spherical geometry.  Their result for a Janus rod is
\begin{equation}
U = \frac{1}{4 \sigma} \left( \frac{R}{L_{1/2}} \right) \ln \left( \frac{L_{1/2}}{4R} \right) ( \mu_+ i_- - \mu_- i_+) = f(L_{1/2},R) \frac{\varepsilon \zeta}{\eta} \frac{i}{\sigma},
\label{V_Golest}
\end{equation}
where $R$ is the radius of the rod, $L_{1/2}$ is its half-length, and subscripts denote the forward (+) and backward ($-$) ends of the rod.  For the case of a uniform mobility ($\mu_+ = \mu_- = \varepsilon \zeta / \eta$) and piecewise uniform current density ($i_+ = - i_- = i$), this equation is identical to the relation of \citet{paxton_catalytically_2006}, except for the geometrical corrections contained in the function $f(L_{1/2},R)$.  In the definition of electrophoretic mobility, $\varepsilon$ is the permittivity of the solution (assumed to be constant), $\zeta$ is the zeta potential of the rod which quantifies its surface charge, and $\eta$ is the viscosity of the solution.  Equation (\ref{V_Golest}) is also applicable, with some modification, to autonomous diffusiophoresis and thermophoresis, which are respectively driven by concentration and temperature gradients generated by the particle.\cite{golestanian_designing_2007}

There have been several recent efforts to gain a more quantitative understanding of the physics driving the motion of bimetallic particles.  In 2010 we took an initial step in this direction, modeling the electrochemical redox reactions with piecewise constant proton fluxes, and assuming a constant, negative electrophoretic mobility for the particle.\cite{moran_locomotion_2010}  With this model, we were able to obtain realistic values of the swimming speed given previously-measured values for the zeta potential and reaction rate.\cite{paxton_catalytically_2006,dougherty_zeta_2008}  In a later work, we provided a more comprehensive and realistic model of the system, using Frumkin-corrected Butler-Volmer kinetics to accurately and self-consistently represent the electrochemical reactions taking place on the rod surface.\cite{moran_electrokinetic_2011}  In both studies, we supplemented our results with scaling arguments based on the governing equations, which capture many of the important dependences of speed on various system parameters.  Our initial scaling result for the swimming speed was
\begin{equation}
U \propto \frac{\varepsilon \zeta}{\eta} \frac{F h \lambda_D j_+}{\varepsilon D_+},
\label{JFMscaling}
\end{equation}
where $F$ is Faraday's constant, $h$ is the length of the rod, $\lambda_D$ is the Debye length which quantifies the thickness of the diffuse screening layer of ions surrounding the particle, $D_+$ is the diffusivity of protons, and $j_+$ is the reaction-driven proton flux.  The proton flux (effectively, the reaction rate) is related to the current density due to the reaction by Faraday's law through the proton valence, $z_+ = 1$, $i = z_+ F j_+$.  Equation (\ref{JFMscaling}) implies a linear proportionality between the speed of the rod and the current density, in agreement with Eqs. (\ref{HSPax}) and (\ref{V_Golest}).  We note the similarity between our scaling relation, (\ref{JFMscaling}), and both Eqs. (\ref{HSPax}) and (\ref{V_Golest}), where the characteristic electric field $E_0 = F h \lambda_D j_+ / \varepsilon D_+$.

Our previous work demonstrated that this velocity depends on the density of the space charge induced by the reactions and the surface charge, as well as the electric field that forms due to the reaction-induced space charge.  To reflect the importance of the reactions and charge distributions in generating this self-propelled motion, we called the autonomous swimming mechanism \textit{reaction induced charge auto-electrophoresis} (RICA).\cite{moran_locomotion_2010,moran_electrokinetic_2011}  The simulations showed strong agreement with the predictions of the scaling analysis.  However, this analysis did not take into account any nonreactive electrolytes that are often present in real systems, although it did implicitly predict an inverse relationship between swimming speed and the square root of background ionic strength (through the direct dependence on Debye length).

Sabass and Seifert presented an analytical and computational study of spherical bimetallic particles that specifically accounts for the presence of a nonreactive salt.\cite{sabass_nonlinear_2012}  They derived an approximate analytical result for the swimming speed and predicted the swimming speed as a function of peroxide concentration both analytically and numerically.  Their analytical result (their Eq. 33) predicts a quadratic dependence of swimming speed on Debye length (equivalently, an inverse dependence on solution ionic strength).  However, they did not calculate the swimming speed as a function of salt concentration or conductivity using either their analytical or numerical models.  They also did not discuss the physical reasons why the conductivity exerts such influence on the swimming speed.

In this paper, we present simulations of bimetallic rod-shaped particles in hydrogen peroxide in the presence of several non-reacting electrolytes.  The simulations solve the full Poisson-Nernst-Planck-Stokes equations with multiple ionic species, a cylindrical particle in an infinite fluid, and nonlinear Butler-Volmer boundary conditions to represent the electrochemical reactions.  The model also accounts for the presence of dissolved carbon dioxide in the form of carbonic acid.  We use three different monovalent salts (potassium chloride (KCl), lithium nitrate (LiNO$_3$), and sodium nitrate (NaNO$_3$)) to vary the solution conductivity and show the differences in results for each electrolyte.  We also derive and validate scaling analyses which predict the dependence of swimming speed and Stern voltage on ionic strength.  The goal of this work is to understand the mechanism that causes the motor's swimming speed to decrease with increasing electrolyte strength.  

%% file: theory.tex

We consider a cylindrical particle, 2 microns in length and 300 nanometers in diameter, suspended in an infinite aqueous solution containing hydrogen peroxide (H$_2$O$_2$), protons (H$^+$), hydroxide ions (OH$^-$), bicarbonate ions (HCO$_3^-$), and one of three nonreactive background electrolytes, which are assumed to dissociate completely: potassium chloride (KCl), sodium nitrate (NaNO$_3$), and lithium nitrate (LiNO$_3$).

We wish to account for all possible contributions to solution conductivity, including the carbonic acid that is often present in aqueous solutions due to the dissolution of atmospheric carbon dioxide.  Carbon dioxide dissolves in water, forming carbonic acid (H$_2$CO$_3$), which can then dissociate twice, forming a bicarbonate ion (HCO$_3^-$) and then a carbonate ion (CO$_3^{2-}$).  It is well-known that the equilibrium state of the system is determined by the partial pressure of CO$_2$ above the solution.\cite{stumm1981aquatic}  Solutions using (for example) MinEQL+ software of the acid dissociation equilibrium equations (shown in the supplemental section), along with the bulk electroneutrality condition (with $z_k$ and $c_k$ as the valence and concentration of ion $k$),
\begin{equation}
\sum_k z_k c_{k,\infty} = 0,
\label{neutrality}
\end{equation}
reveal that for a pH in the typical range for dissolved atmospheric carbonic acid, the equilibrium concentrations of OH$^-$ and CO$_3^{2-}$ are lower than that of HCO$_3^-$ by nearly three orders of magnitude.  Here the subscript $\infty$ indicates the value in the bulk.  Thus, we ignore the presence of OH$^-$ and CO$_3^{2-}$ and assume a binary electrolyte composed of H$^+$ and HCO$_3^-$.  We assume that the partial pressure of CO$_2$ above the solution is $5.6 \times 10^{-5}$ atm, yielding bulk concentrations of H$^+$ and HCO$_3^-$ of $c_{\pm,\infty} = 9 \times 10^{-7}$~mol/L, implying a pH of 6.05.  We also assume that the dissociation of hydrogen peroxide into H$^+$ and HO$_2^-$ is negligible, since it was previously shown that peroxide does not dissociate significantly enough to affect the pH of the solution or the swimming speed.\cite{sabass_nonlinear_2012}

The geometry is axisymmetric and therefore we use a cylindrical coordinate system where all variables are independent of the azimuthal angle, $\theta$.  The simulations are conducted in a two-dimensional slice of the three-dimensional domain with an area $100 \times 100$~$\mu$m$^2$.  The rod is positioned with its centroid at $(r,z) = (0,0)$, at the midpoint of the domain's axis of symmetry.

\subsection{Governing Equations and Scaling Analysis}
Following our previous work,\cite{moran_locomotion_2010,moran_electrokinetic_2011} we apply the Poisson-Nernst-Planck-Stokes system of equations to this problem.  In the dilute solution limit, the concentration distributions of all species obey the dimensionless advection-diffusion equation,
\begin{equation}
Ra_e \left( \tilde{\mathbf{u}} \cdot \tilde{\nabla} \tilde{c}_k \right) = \tilde{\nabla}^2 \tilde{c}_k - \beta_k \tilde{\nabla } \cdot \left( \tilde{c}_k \tilde{\mathbf{E}} \right),
\label{AD}
\end{equation}
where $\tilde{\mathbf{u}}$ is the fluid velocity normalized by the electroviscous velocity $U_{ev}$, $\tilde{c}_k$ is the concentration of species $k$ normalized by the background proton and bicarbonate ion concentration, $c_{\pm,\infty}$, $\tilde{\mathbf{E}} = -\tilde{\nabla} \tilde{\phi}$ is the electric field normalized by a characteristic electric field $E_0$,  $Ra_e$ is the electric Rayleigh number,
\begin{equation}
Ra_e = \frac{U_{ev} a }{D_+},
\label{Raedef}
\end{equation}
$D_+$ is the diffusivity of protons, $a$ is a length scale over which the tangential electric field is significant, and $\beta_k$ is a dimensionless parameter that quantifies the relative importance of electromigration and diffusion of ion $k$ given as,
\begin{equation}
\beta_k = \frac{z_k F E_0 a}{RT}.
\label{betadef}
\end{equation}
In this case, all ions are monovalent, and the parameter $\beta$ therefore has the same value for every ion. Throughout the paper, dimensionless variables are indicated with a tilde, while dimensional variables and constants have no tilde.  For oxygen and hydrogen peroxide, which are uncharged, the electromigration term is omitted.

The concentration distributions are coupled to the electrostatic potential distribution through Poisson's equation,
\begin{equation}
- \frac{\varepsilon E_0}{F a c_{\pm,\infty}} \tilde{\nabla}^2 \tilde{\phi} = \tilde{\rho}_e = \sum_k z_k \tilde{c}_k.
\label{poisson}
\end{equation}
Here $\varepsilon$ is the permittivity of the solution, $c_{\pm,\infty}$ is the bulk ion concentration, $\tilde{\phi}$ is the dimensionless electric potential normalized by $E_0 a$, and the summation is carried out over all ionic species.  The fluid flow is described by the incompressible continuity and Stokes equations:
\begin{equation}
\tilde{\nabla} \cdot \tilde{\mathbf{u}} = 0,
\label{COM}
\end{equation}
\begin{equation}
0 = \frac{1}{Re} \left( - \tilde{\nabla} \tilde{p} + \tilde{\nabla}^2 \tilde{\mathbf{u}} + \tilde{\rho}_e \tilde{\mathbf{E}} \right).
\label{stokes}
\end{equation}
Here $Re = \rho U_{ev} d / \eta$ is the Reynolds number (which is never larger than 10$^{-4}$ in the simulations considered here), $\tilde{p}$ is the pressure normalized by $\eta U_{ev}/d$,  $d$ is a viscous length scale, and $\tilde{\rho}_e \tilde{\mathbf{E}}$ is the electrical body force resulting from the coupling of free charge in the solution and the self-generated electric field.  Equations (\ref{AD}) and (\ref{poisson})-(\ref{stokes}) constitute a coupled, nonlinear system that is difficult to solve in general.  Approximate versions of these equations have been solved analytically to study the self-propulsion of a spherical cell,\cite{lammert_ion_1996} the autonomous fluid circulation near metallic disk electrodes in a peroxide solution,\cite{kline_catalytically_2006} and recently electrokinetic self-propulsion of synthetic particles similar to the case considered here.\cite{yariv_electrokinetic_2011,sabass_nonlinear_2012}  Generally, obtaining analytical solutions to these equations requires one to make simplifying assumptions, e.g. that the reactions cause small perturbations of the field variables from the equilibrium state,\cite{kline_catalytically_2006,yariv_electrokinetic_2011} or that the electrical double layer (EDL) surrounding the particle is negligibly thin compared to the size of the particle.\cite{yariv_electrokinetic_2011,sabass_nonlinear_2012}  We wish to describe in rigorous detail the physical phenomena occurring at the rod/solution interface, and we accordingly make no simplifying assumptions about the size of the EDL or the concentration perturbations. This allows us to study the system for a wide range of parameter values, and obtain numerical solutions to the full nonlinear equations to account for the cylindrical geometry and incorporate all of the important physical phenomena leading to self-propulsion.

The natural velocity scaling in this system is the electroviscous velocity, $U_{ev} = \rho_{e,0} E_0 / (\eta / d^2)$, which was originally introduced by \citet{hoburg_internal_1976} and reflects the balance of viscous and electrical body forces in the system.  We expect that the speed of the rod scales with $U_{ev}$.  In our previous work we derived the general scaling relation for a charged rod with zeta potential $\zeta$ and area-averaged proton flux $j_+$ given as
\begin{equation}
U_{ev} \propto \frac{F L a d^2}{\lambda_D^2 \eta D_+} j_+ \zeta,
\label{UevJFM}
\end{equation}
where $L$ is a length scale for the charge density distribution.  Previously, we assumed that $L$ and $d$ both scale with the Debye length, while $a$ was proportional to the length of the rod, $h$.\cite{moran_locomotion_2010,moran_electrokinetic_2011}  Although we have confirmed that the tangential electric field does increase in magnitude with increasing $h$ (assuming the particle is suspended in an infinite medium and neglecting the role of the particle mass), as shown in the supplementary section, here we are specifically interested in changes in electric field due to changes in solution conductivity.  According to Ohm's law, $i = \sigma E$, the electric field should scale inversely with solution conductivity.  Solution conductivity scales approximately with ionic strength, $I$, and therefore we would expect $E_0 \propto 1 / I$.  Ionic strength is related to the Debye length according to
\begin{equation}
\lambda_D^2 \equiv \frac{\varepsilon RT}{2 F^2 I}.
\end{equation}
From this definition and the inverse relationship between electric field and ionic strength, we see that the electric field can scale with the square of Debye length, $E_0 \propto \lambda_D^2$.\cite{MoranPhD13}  Note that we are not claiming that the electric field scales with the physical thickness of the EDL.  Instead, we concur with previous work\cite{paxton_catalytically_2006} that the electric field scales inversely with the conductivity of the solution, through Ohm's law, which we express in terms of the definition of the Debye thickness.  Making this substitution into equation (\ref{UevJFM}), the scaling result for the swimming speed becomes
\begin{equation}
U_{ev} \propto \frac{F \lambda_D^2}{\eta D_+} j_+ \zeta \propto \frac{\varepsilon \zeta}{\eta} \frac{RT}{FD_+ I} j_+.
\label{scaling}
\end{equation}
Again, this relation resembles the Helmholtz-Smoluchowski-like expression, Eq. (\ref{HSPax}), except with the effective electric field given by $E_0 \propto RT j_+ / FD_+ I$, and the equation is stated in terms of reaction rate $j_+$ instead of current density $i$.  This equation predicts that the swimming speed should scale quadratically with Debye length (or inversely with ionic strength).  A quadratic relationship between speed and Debye length was also asymptotically derived by Sabass and Seifert. \cite{sabass_nonlinear_2012}  The prediction of an inverse dependence on conductivity was also made by \citet{paxton_catalytically_2006} and by Golestanian, Liverpool and Ajdari.\cite{golestanian_designing_2007}  In all three cases, the predicted form for the swimming speed is proportional to the Helmholtz-Smoluchowski expression, except with different forms for the electric field.  In addition, Golestanian's formula and our simulations account for the cylindrical geometry of the particle.

\subsection{Boundary Conditions}
The simulations are conducted in the reference frame of a stationary rod.  We apply the no-slip condition at the rod surface,
\begin{equation}
\tilde{\mathbf{u}} = \mathbf{0}.
\end{equation}
At the domain boundary far from the rod, we prescribe vanishing viscous stress:
\begin{equation}
\left[ \tilde{\nabla} \tilde{\mathbf{u}} + \left( \tilde{\nabla} \tilde{\mathbf{u}} \right)^T \right] = 0.
\end{equation}
Here $\tilde{\nabla} \tilde{\mathbf{u}}$ is the velocity gradient tensor and the superscript $T$ denotes the transpose.  This boundary condition effectively enforces a slip condition at the outer boundary, approximating an infinite medium.  We evaluate the average fluid speed in the axial direction along this boundary to determine the swimming speed of the rod.

Since anions do not react, their boundary condition is zero flux at the rod surface,
\begin{equation}
\mathbf{n} \cdot \tilde{\mathbf{j}}_- = 0,
\end{equation}
where $\mathbf{n}$ is the outward normal vector pointing into the fluid and the dimensionless flux of ion $k$ is defined as $\tilde{\mathbf{j}}_k = -\tilde{\nabla} \tilde{c}_k + \beta_k \tilde{c}_k \tilde{\mathbf{E}} + Ra_e \tilde{c}_k \tilde{\mathbf{u}}.$  The electrochemical reactions generate fluxes of protons leaving the anode surface and entering the cathode surface.  These reactions are represented (dimensionally) in the model by
\begin{equation}
\mathbf{n} \cdot \mathbf{j}_+ = \left\{
                                          \begin{array}{ll}
                                          j_{+,a} = K_{ox} c_{\mathrm{H}_2 \mathrm{O}_2} \exp \left[ \frac{(1-\alpha) m F \Delta \phi_S}{RT} \right], \textrm{   }0 < z < 1 \textrm{ }\mu\textrm{m},\\
                                          j_{+,c} = K_{red} c_{\mathrm{H}_2 \mathrm{O}_2} c_+^2 \exp \left[ - \frac{\alpha m F \Delta \phi_S}{RT} \right], \textrm{  } - 1 \textrm{ }\mu\textrm{m} < z < 0
                                          \end{array}
                                          \right.
\label{fluxbc}
\end{equation}
where the subscript + denotes protons, the subscript $a$ indicates the anodic flux due to the peroxide oxidation reaction, and the subscript $c$ indicates cathodic flux for the peroxide reduction reaction.  Positive values of the axial coordinate $z$ indicate the anode side of the rod (typically Pt), and negative values indicate the cathode (typically Au) side.  Here, $K_{ox}$ and $K_{red}$ are the rate constants for peroxide oxidation and reduction, $\alpha$ is a dimensionless parameter between 0 and 1 (set here to 0.5) that quantifies the asymmetry of the energy barrier for the reaction, $m$ is the number of electrons transferred in the electrochemical reaction ($m=2$ for both reactions considered here), and $\Delta \phi_S$ is the voltage across the compact Stern layer of adsorbed species on the surface of the rod.\cite{bazant_current-voltage_2005}

The expressions for the proton fluxes $j_+$ on the rod segments are given by Butler-Volmer equations with Frumkin's correction,\cite{frumkin_hydrogen_1933} and reflect the dependence of the kinetics of the electrochemical reactions on the local reactant concentrations and on the voltage across the compact Stern layer.\cite{moran_electrokinetic_2011,frumkin_hydrogen_1933,delahay_double_1965,bard_electrochemical_2000,bazant_current-voltage_2005}  In equation (\ref{fluxbc}), we have implemented the Tafel approximation, meaning that the reactions are assumed to proceed in one direction only and the backward components of each reaction are considered negligible.  For a derivation of Eq. (\ref{fluxbc}) starting from the full Butler-Volmer equation, the reader is referred to our previous paper.\cite{moran_electrokinetic_2011}

The reaction rates for species other than protons are related to the proton fluxes according to the stoichiometry of the reactions.  On the anode, one peroxide molecule is consumed for every two protons released into the solution:
\begin{equation}
\mathbf{n} \cdot \mathbf{j}_{\textrm{H}_2\textrm{O}_2,a} = - \frac{j_{+,a}}{2},
\end{equation}
and one oxygen molecule is generated for every two protons generated:
\begin{equation}
\mathbf{n} \cdot \mathbf{j}_{\textrm{O}_2,a} = \frac{{j}_{+,a}}{2}.
\end{equation}
On the cathode, one peroxide molecule is consumed for every two protons consumed:
\begin{equation}
\mathbf{n} \cdot \mathbf{j}_{\textrm{H}_2\textrm{O}_2,c} = \frac{j_{+,c}}{2}.
\end{equation}
It has been suggested by Wang \textit{et al.},\cite{wang_bipolar_2006} among others, that the four-electron reduction of O$_2$ may also occur on the cathode end, perhaps even as the dominant reaction.  While this is possible, O$_2$ reduction is not likely the dominant reduction reaction on the cathode side, since our previous work shows that the rods move faster in solutions purged of O$_2$ and slower in O$_2$-rich solutions.\cite{calvo-marzal_electrochemically-triggered_2009}  We therefore assume that peroxide reduction is the only reaction occurring on cathode, and that oxygen is nonreactive on the cathode end:
\begin{equation}
\mathbf{n} \cdot \mathbf{j}_{\textrm{O}_2,c} = 0.
\end{equation}
Far from the rod, the concentrations of all chemical species approach their bulk values,
\begin{equation}
\tilde{c}_k \rightarrow 1 \textrm{ as } \left| \tilde{\mathbf{r}} \right| \rightarrow \infty.
\end{equation}
The boundary conditions for proton flux can also be stated in dimensionless form.  Since the proton flux takes a different form on the anode and cathode, here the Damk\"{o}hler number is defined differently for each metal.  The Damk\"{o}hler numbers can be defined in terms of the reaction kinetic expressions, equation (\ref{fluxbc}).  On the anode, the dimensionless boundary condition takes the form
\begin{equation}
\mathbf{n} \cdot \tilde{\mathbf{j}}_+ = Da_{anode} \tilde{c}_{\textrm{H}_2 \textrm{O}_2},
\end{equation}
where
\begin{equation}
Da_{anode} = \frac{K_{ox} a c_{\textrm{H}_2 \textrm{O}_2,\infty}}{D_+ c_{+,\infty}}.
\label{Da_anode_def}
\end{equation}
On the cathode, the dimensionless boundary condition reads
\begin{equation}
\mathbf{n} \cdot \tilde{\mathbf{j}}_+ = Da_{cathode} \tilde{c}_{\textrm{H}_2 \textrm{O}_2} \tilde{c}_+^2,
\end{equation}
where
\begin{equation}
Da_{cathode} = \frac{K_{red} a c_{\textrm{H}_2 \textrm{O}_2,\infty} c_{+,\infty}}{D_+}.
\label{Da_cathode_def}
\end{equation}
Although the Damk\"{o}hler numbers are defined differently on the anode and cathode, the rate constants in each definition have different units, so that the Damk\"{o}hler number is dimensionless in each case.  In stating the definitions of the Damk\"{o}hler numbers we have ignored the exponential terms in the kinetic expressions (i.e., we have assumed these terms to be equal to unity).  Although these terms could be included, they would not significantly alter the magnitude of the Damk\"{o}hler numbers.  In this work, the exponential terms range in magnitude from 0.97 to 1.03 in all cases studied.

In general, the flux expressions for the anode and cathode are not equal at the junction between them, $z = 0$.  To avoid unphysical discontinuities in the reaction flux and flux gradient, we multiply the flux profile along the length of the rod by a dimensionless sigmoidal weighting function, $\xi (z)$, defined as
\begin{equation}
\xi (z) = \left| \frac{2}{1 + e^{- \gamma z}} - 1 \right|,
\end{equation}
where $\gamma = 10^7$~m$^{-1}$ and $z$ is evaluated in meters.  The function $\xi$ is defined to be roughly equal to 1 at the end of the anode segment, 1 at the end of the cathode segment, and zero at the anode/cathode boundary.  The use of this weighting function reflects a diffuse interface which would result in reduced density of available reaction sites and reaction rate near the junction between anode and cathode.

The surface of the rod is theorized to contain an immobile layer of charged and uncharged adsorbed species, often referred to as the \textit{Stern layer}.  Together with the diffuse layer of ions in the solution adjacent to the rod, these two layers constitute the electrical double layer (EDL).  According to the Stern model of the EDL, the immobile (Stern) layer acts as a linear capacitor in series with the diffuse layer.\cite{bard_electrochemical_2000,bazant_current-voltage_2005}  The electric potential gradient is extrapolated across the Stern layer, from the outer Helmholtz plane to the metal.  Thus, the Stern voltage is linearly related to the normal electric field at the rod surface.  Following the Stern model, we treat this layer as a linear capacitor which leads to the (dimensional) mixed boundary condition\cite{bard_electrochemical_2000,bazant_current-voltage_2005,moran_electrokinetic_2011}
\begin{equation}
\Phi_{rod} + \lambda_S \left( \mathbf{n} \cdot \nabla \phi \right)_{\textrm{OHP}} = \phi_{\textrm{OHP}} \equiv \zeta,
\label{potentialBC}
\end{equation}
where $\Phi_{rod}$ is the electrical potential of the interior of the rod with respect to the bulk solution, $\lambda_S$ is an effective thickness of the Stern layer (set here to 2 {\AA} for all cases), and the subscript OHP indicates that the quantity is evaluated at the outer edge of the Stern layer, often termed the outer Helmholtz plane (OHP).\cite{bard_electrochemical_2000}  Since the rod is conducting, the potential $\Phi_{rod}$ is assumed uniform everywhere in its interior.  The zeta potential $\zeta$ is defined here as the potential at the OHP versus the bulk solution, and in general varies with position on the rod surface.  The voltage across the Stern layer is generally defined as the internal rod potential minus the potential at the OHP, i.e. $\Delta \phi_S \equiv \Phi_{rod} - \zeta (z)$, and therefor depends on the normal electric field at the OHP through the above boundary condition on the potential, (\ref{potentialBC}).

Far from the rod, the electric potential approaches zero,
\begin{equation}
\tilde{\phi} \rightarrow 0\textrm{ as }\left| \tilde{\mathbf{r}} \right| \rightarrow \infty.
\end{equation}

\subsection{Current Conservation}
At steady state, the total charge in the rod must be conserved, implying that the net current into or out of the rod must be zero.  We require that
\begin{equation}
\int_{anode} j_{+,a} dA = - \int_{cathode} j_{+,c} dA \equiv J
\label{currcons}
\end{equation}
at steady state, where the reaction fluxes $j_{+,a}$ and $j_{+,c}$ are given by (\ref{fluxbc}).  The system of equations (\ref{AD})-(\ref{stokes}) is solved concurrently and is closed by iterative determination of the rod potential $\Phi_{rod}$ that produces reaction fluxes that satisfy (\ref{currcons}).  The value of $\Phi_{rod}$ directly affects the reaction rates on both the anode and cathode.  On the cathode, a more negative rod potential would result in a more negative zeta potential, which means that the surface attracts more protons electrostatically to the surface to screen the surface charge.  The elevated proton concentration results in faster reaction rates on the cathode, according to (\ref{fluxbc}).  On the anode, a more negative potential decreases the reaction rate, since this would alter the overpotential bias to favor reduction more and oxidation less.  The value of $\Phi_{rod}$ that satisfies (\ref{currcons}) is observed to vary with salt concentration.  Table \ref{salttab} shows the values and units of the constants used in the simulations.\cite{lide_crc_2004}  The ion mobilities $\nu_k$ are determined from the Nernst-Einstein relation, $D_k = \nu_k RT$.

\input{salttab.tex} 

%% file: salttab.tex
\begin{table}

\begin{minipage}[c]{13cm}
\centering
\begin{tabular}{l l r}
\textbf{Constant} \hspace{0.2cm} & \hspace{0.2cm} \textbf{Description} \hspace {0.2cm} & \hspace{0.4cm} \textbf{Value} \hspace{0.2cm} \\
\hline \hline
$K_{ox}$ & Oxidation rate constant, anode & 2.2 $\times 10^{-7}$ m s$^{-1}$  \\
$K_{red}$ & Reduction rate constant, cathode & 1 m$^7$ s$^{-1}$ mol$^{-2}$ \\
$D_+$ & Diffusivity, protons & 9.311 $\times 10^{-9}$ m$^2$ s$^{-1}$ \\
$D_-$ & Diffusivity, bicarbonate ions & 5.273 $\times 10^{-9}$ m$^2$ s$^{-1}$ \\
$D_{\textrm{H}_2\textrm{O}_2}$ & Diffusivity, hydrogen peroxide & 6.6 $\times 10^{-10}$ m$^2$ s$^{-1}$ \\
$D_{\textrm{O}_2}$ & Diffusivity, molecular oxygen & 2 $\times 10^{-9}$ m$^2$ s$^{-1}$ \\
$D_{\textrm{Li}^+}$ & Diffusivity, lithium ion & 1.029 $\times 10^{-9}$ m$^2$ s$^{-1}$ \\
$D_{\textrm{Na}^+}$ & Diffusivity, sodium ion & 1.334 $\times 10^{-9}$ m$^2$ s$^{-1}$ \\
$D_{\textrm{K}^+}$ & Diffusivity, potassium ion & 1.957 $\times 10^{-9}$ m$^2$ s$^{-1}$ \\
$D_{\textrm{NO}_3^-}$ & Diffusivity, nitrate ion & 1.902 $\times 10^{-9}$ m$^2$ s$^{-1}$ \\
$D_{\textrm{Cl}^-}$ & Diffusivity, chloride ion & 2.032 $\times 10^{-9}$ m$^2$ s$^{-1}$ \\
$\eta$ & Solution viscosity & 8.9 $\times 10^{-4}$ Pa s \\
$\rho$ & Solution density & 998 kg m$^{-3}$ \\
$C_{\pm,\infty}$ & Bulk concentration, protons and bicarbonate ions & $9 \times 10^{-7}$ mol L$^{-1}$ \\
$C_{\textrm{O}_2,\infty}$ & Bulk concentration, molecular oxygen & $0.2 \times 10^{-3}$ mol L$^{-1}$ \\
$C_{\textrm{H}_2\textrm{O}_2,\infty}$ & Bulk concentration, hydrogen peroxide & 1.11 mol L$^{-1}$ \\
$\varepsilon_r$ & Solution dielectric constant & 78.4 \\
$\lambda_S$ & Effective Stern layer thickness & 0.2 nm \\
\hline
\end{tabular}
\end{minipage}
\caption{Relevant constants used in the simulations.  The bulk electrolyte concentration is not shown because it is varied from 56.4 to 820 $\mu$mol/L throughout the work.  The rate constants $K_{ox}$ and $K_{red}$ are fitting parameters and are chosen to yield a swimming speed approximately equal to that observed by \citet{paxton_catalytically_2006} at a conductivity of 8.8~$\mu$S/cm.}
\label{salttab}

\end{table}

%% file: results.tex

In this section, we present the results of simulations with three different monovalent electrolytes at different concentrations.  In all cases, the bulk hydrogen peroxide concentration is set at 1.11~mol/L (3.7 wt. \%) to facilitate comparison of this work with that of \citet{paxton_catalytically_2006}  By varying electrolyte concentration, we observe the variation in the distributions of proton concentration, electric potential, electric field, and velocity, and thereby determine the variation in swimming speed with solution conductivity.  We compare our numerical calculations to previous analytical and experimental results.

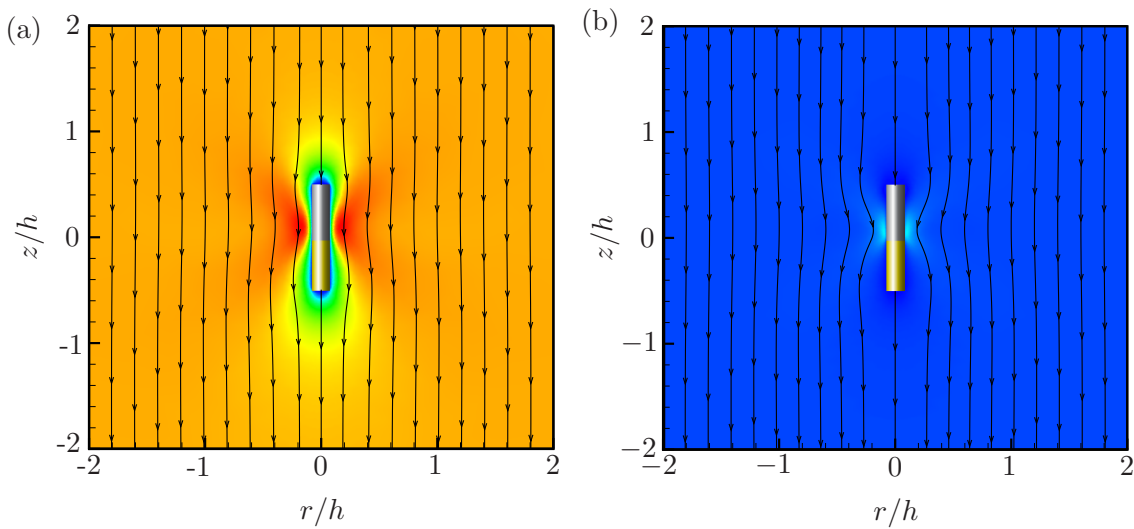
\begin{figure}
\centering
\input{fig1a.tex}
\input{fig1b.tex}
\caption{Simulation-generated plots of velocity field and flow streamlines around a Pt/Au rod.  The Pt (anode, top) end is colored silver, and the Au (cathode, bottom) end is colored gold.  Panel a shows the case for no added salt [H$_+$] = [HCO$_3^-$] = 0.9~$\mu$mol/L.  Panel b shows the case for added background electrolyte with concentration set to [K$^+$] = [Cl$^-$] = 56.4 $\mu$M, yielding a conductivity of 8.8 $\mu$S/cm.  To facilitate comparison, the color scale is the same for both figures.  Blue corresponds to low, green and yellow to moderate, and red to high fluid velocity magnitude.}
\label{fig1}
\end{figure}

To illustrate the qualitative differences caused by varying electrolyte strength, we show in Fig. \ref{fig1} contour plots of fluid velocity magnitude for cases with (a) water and peroxide in equilibrium with atmospheric carbon dioxide and (b) the same case with added KCl.  Streamlines of fluid flow are overlaid onto each plot.  Figure \ref{fig1} (a) visualizes a simple case with only peroxide, oxygen, protons, and bicarbonate ions (the latter two at a concentration of 0.9 $\mu$mol/L, resulting in a solution conductivity of 0.35 $\mu$S/cm) and no additional salt.  In Fig. \ref{fig1} (b), KCl has been added at a concentration of 56.4 $\mu$mol/L (in addition to the 0.9 $\mu$mol/L protons and bicarbonate) such that the bulk conductivity is 8.8 $\mu$S/cm, which is the lowest conductivity reported by \citet{paxton_catalytically_2006}  Considering that the color scales are the same in these two figures, it is clear that the speed of the rod is significantly reduced in the salt case, (b).

In both plots in Fig. \ref{fig1}, a region of high fluid velocity magnitude is clearly visible near the equator of the rod ($z = 0$).  These high-speed regions appear because of the extremely strong charge density and electric fields near the rod.  The charge density in the EDL, which arises to screen the surface charge, scales inversely with the square of Debye length.\cite{moran_electrokinetic_2011}  As salt is added to the solution, the diffuse layer shrinks, significantly increasing the magnitude of the charge density in the diffuse layer.  This charge density couples with the $z$-direction electric field, which is especially strong in magnitude in this region due to the large gradient in charge density, to produce strong electrical body forces in the vicinity of the rod.  The body forces are strongest near the anode/cathode junction, leading to the above-average fluid speed near the junction.  As Fig. \ref{fig1} shows, the region of high velocity magnitude becomes more prominent relative to the background flow as salt concentration is increased.

Near the center of Fig. \ref{fig1} (b), the flow streamlines bend noticeably inward toward the rod.  This bending occurs due to the aforementioned increase in flow speed near the rod surface.  The contraction of the streamlines near the rod surface is a natural consequence of mass conservation, and satisfies the requirement for incompressible flows that the volume flow rate between two adjacent streamlines must be constant.\cite{cengel_fluid_2010}  This effect is more noticeable at high conductivities because the region of high velocity becomes more prominent (i.e., larger speeds compared to the surrounding fluid) as salt concentration is increased, as discussed above.

\begin{figure}
\centering
\input{fig2a.tex}
\input{fig2b.tex}
\caption{(a) Axial swimming speed for bimetallic rods vs. solution conductivity in 3.7 \% H$_2$O$_2$ for experiments, theory, and simulations.  Open symbols indicate the current computational work, with KCl ($\bigcirc$), NaNO$_3$ ($\square$) and LiNO$_3$ ($\lozenge$) as the electrolyte.  Experimental data for NaNO3 ($\blacksquare$) and LiNO3 ($\blacktriangle$) is taken from \citet[\,]{paxton_catalytically_2006}  ($\times$) symbols indicate data from the analytical formula derived by Golestanian, Liverpool, and Ajdari.\cite{golestanian_designing_2007}  The dashed line is a fit curve of the form $U \propto \sigma^{-1}$.  (b) Simulation data plotted versus ionic strength.  Here, the dashed scaling line is a fit of the form $U \propto C^{-1}$.}
\label{fig2}
\end{figure}
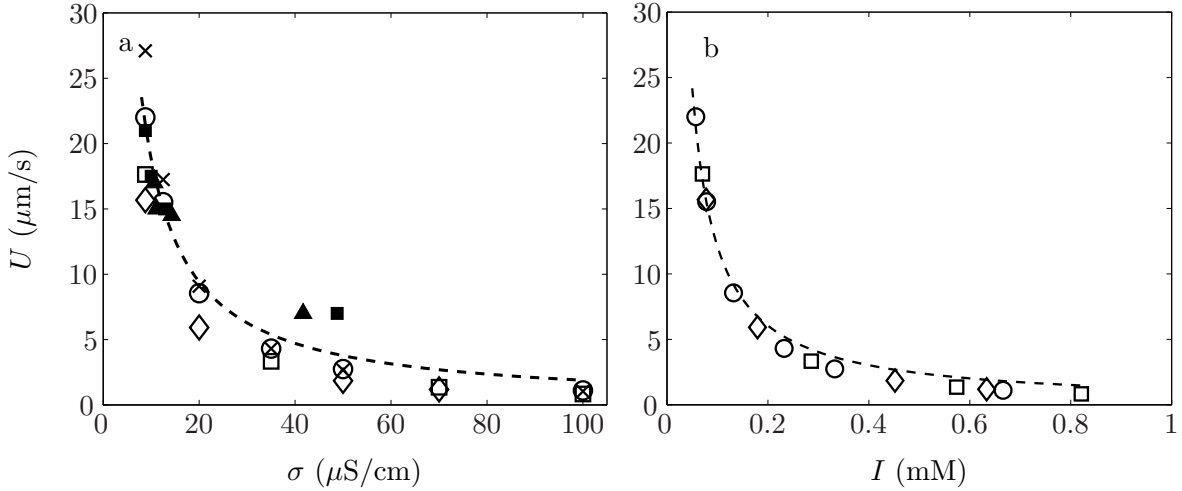

Figure \ref{fig2} (a) shows the motor swimming speed versus solution conductivity for the simulations with added KCl, NaNO$_3$ and LiNO$_3$, the experimental data of \citet{paxton_catalytically_2006} for NaNO$_3$ and LiNO$_3$, Golestanian's formula\cite{golestanian_designing_2007} with the values of the phoretic mobility and surface current density determined from our current simulations.  The solution conductivity at infinite dilution is calculated from $\sigma = \sum_k z_k^2 \Lambda_k c_{k,\infty}$, where $\Lambda_k$ is the molar conductivity of ion $k$, which depends on its mobility.  Golestanian's formula, Eq. (\ref{V_Golest}), requires the surface current density $i$, electrophoretic mobility $\mu_e$, and conductivity $\sigma$ as inputs.  At a given conductivity, we use the area-averaged zeta potential $\bar{\zeta}$ (determined from the simulations) to calculate the electrophoretic mobility as $\varepsilon \bar{\zeta} / \eta$ and use the area-averaged current density out of the anode (also determined from the simulations) for the surface current density.

Our simulations and Paxton's experiments are comparable at low conductivity because the values of the rate constants $K_{ox}$ and $K_{red}$ were chosen to yield identical values for the swimming speed at a conductivity of 8.8 $\mu$S/cm.  Figure \ref{fig2} (a) shows that the simulations reproduce the same trend of speed vs. conductivity that is observed experimentally.  Golestanian's relation, Eq. (\ref{V_Golest}), predicts nearly exactly the same swimming speed as the simulations given the simulation-fed parameters, especially at high conductivity.  The deviations between Golestanian's theory and the simulations at low conductivities is attributed to the assumption inherent in Eq. (\ref{V_Golest}) that the EDL is infinitely thin, which becomes less accurate as conductivity is decreased and the EDL thickness becomes finite.  Note that several of the trends (speed proportional to surface current density, phoretic mobility, and inversely proportional to conductivity) predicted by Golestanian's equation were also predicted by Paxton's equation\cite{paxton_catalytically_2006} and by our scaling relation, Eq. (\ref{scaling}).\cite{moran_locomotion_2010,moran_electrokinetic_2011}  Paxton's and our scaling relations would quantitatively overpredict the velocity, given the current density and zeta potential.  However, by accounting for the non-spherical geometry of the particle, Golestanian's formula is capable of predicting the correct magnitude of the swimming velocity (especially at high conductivities) given the inputs from the simulations, as shown in Fig. \ref{fig2} (a).  For the rod with dimensions considered here, these geometrical corrections are equal to approximately 0.038.

Although the increase in solution conductivity affects the current density, electric field, and ultimately velocity field distributions, it is not clear that the conductivity is the fundamental parameter which controls the swimming speed.  Figure \ref{fig2} (a) shows that at the same value of conductivity, the swimming speeds are generally different for the three salts.  Figure \ref{fig2} (b) shows the same simulation data as Fig. \ref{fig2} (a), plotted instead versus ionic strength.  Ionic strength is defined as $C = (1/2) \sum_k z_k^2 c_{k,\infty}$, and differs from conductivity in that the mobilities of the ions are not considered.  All three data sets in Fig. \ref{fig2} (b) collapse onto the same curve, suggesting that the ion mobilities are not important in determining the swimming speed of the rod.

\begin{figure}
\centering
\input{fig3.tex}
\caption{Swimming speed data plotted versus Debye length, varied by varying the concentration of KCl.  The ($\blacksquare$) symbols indicate simulation data and the solid line is a quadratic fit to the data.  This figure strongly suggests that the quadratic dependence of speed on Debye length (inversely on solution ionic strength) predicted by the scaling result (\ref{scaling}) and derived by \citet{sabass_nonlinear_2012} is valid.}
\label{fig3}
\end{figure}
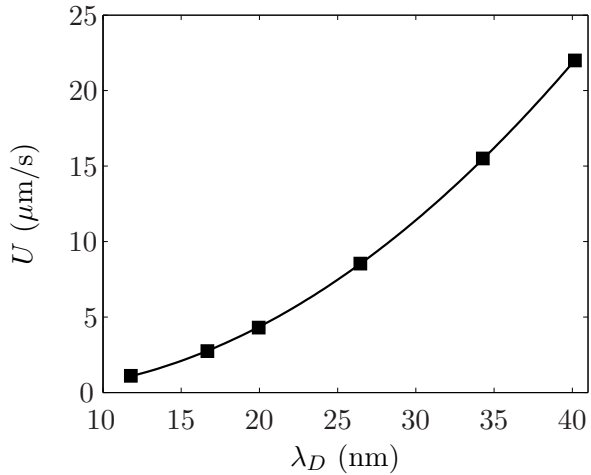

Figure \ref{fig3} shows the swimming speed data as a function of Debye length along with the scaling relation, Eq. (\ref{scaling}).  The near-perfect agreement between the simulation data and the scaling relation (shown as a quadratic fit curve) suggests that the swimming speed is directly related to square of Debye length or, equivalently, the reciprocal of solution ionic strength.\cite{MoranPhD13}

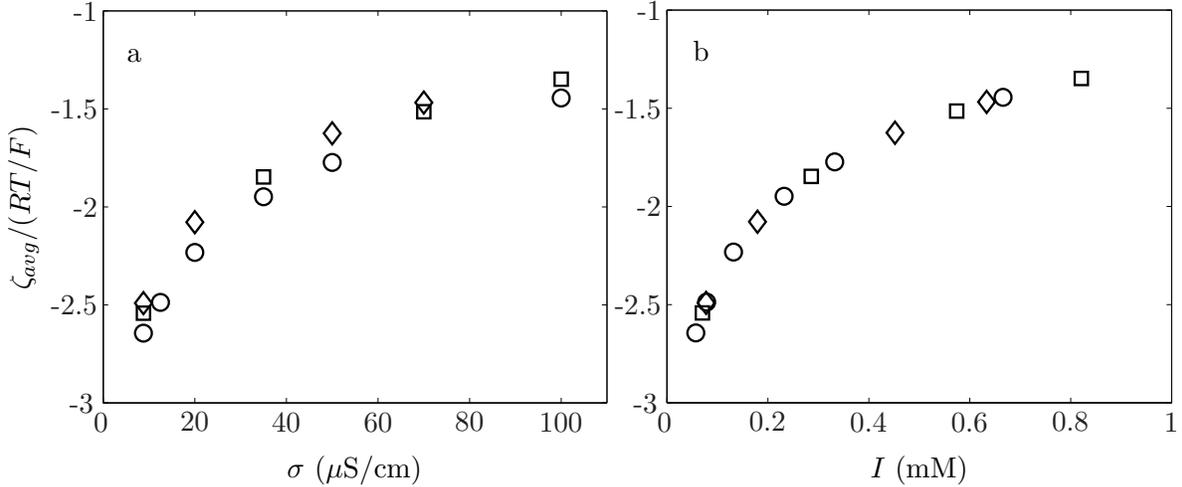
\begin{figure}
\centering
\input{fig4a.tex}
\input{fig4b.tex}
\caption{Area-averaged zeta potential, normalized by the thermal voltage, $RT/F = 25.6$~mV, of the rod versus solution conductivity (a) and ionic strength (b) for three different electrolytes: KCl ($\bigcirc$), NaNO$_3$ ($\square$), and LiNO$_3$ ($\lozenge$).  When plotted versus ionic strength, the data collapse onto a single curve.  Close examination of (a) and Fig. \ref{fig2} (a) confirms that motor velocity is a function of the rod potential.}
\label{fig4}
\end{figure}

Several previous analyses, including our own, predicted a linear relationship (proportional to the Helmholtz-Smoluchowski equation) between the surface charge on a self-electrophoretic swimmer and its swimming speed.  Examples include the work of Lammert, Prost, and Bruinsma,\cite{lammert_ion_1996}, Paxton \textit{et al.},\cite{paxton_motility_2005,paxton_catalytically_2006} Golestanian, Liverpool, and Ajdari,\cite{golestanian_designing_2007} Sabass and Seifert,\cite{sabass_nonlinear_2012} and our previous work.\cite{moran_locomotion_2010,moran_electrokinetic_2011}  Here, we again observe a strong correlation between speed and rod potential. Figs. \ref{fig4} (a) and (b) show the variation in rod zeta potential with conductivity and ionic strength, respectively.  When electrolyte is added to the system, the reaction rates are affected asymmetrically, requiring the rod potential to become less negative in order to satisfy the current conservation constraint expressed by equation (\ref{currcons}).  Thus, a change in electrolyte concentration leads to a change in rod potential.  By the boundary condition (\ref{potentialBC}), this change in rod potential changes the zeta potential.  Figures \ref{fig4} (a) and (b) show that the magnitude of the zeta potential decreases as electrolyte is added to the solution.

\begin{figure}
\centering
\input{fig5.tex}
\caption{Electric field (in the z direction) at 50~nm from the rod surface, $E_{50}$ ($\blacksquare$), and the characteristic electric field magnitude $E^*$($\bullet$), defined in the text, as a function of solution conductivity.  Both electric fields have been normalized by Ohmic electric field at the minimum conductivity, $i/\sigma = 1.32 \times 10^4$~V/m.  Solid lines are power-law fits to the data.  The fit results show that $E_{50} \propto \sigma^{-0.6}$, while $E^* \propto \sigma^{-0.988}$.}
\label{fig5}
\end{figure}
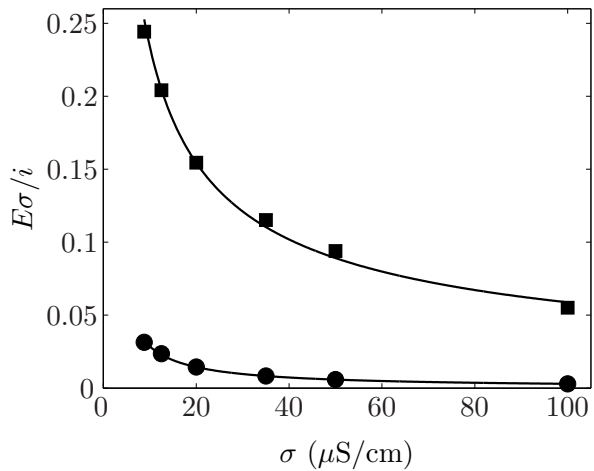

Figure \ref{fig5} shows the $z$-direction electric field $E_{50}$ measured at the junction between the anode and cathode, 50 nm from the rod surface, and the characteristic electric field $E^*$, as a function of solution conductivity.  $E^*$ is defined as the externally applied electric field that would be required to drive conventional electrophoresis of a particle at a speed equal to the measured swimming speed, $U$.  That is,
\begin{equation}
E^* \equiv U \frac{\eta}{\varepsilon \bar{\zeta}}.
\label{Estardef}
\end{equation}
This definition is equivalent to that used in our previous work,\cite{moran_electrokinetic_2011} except here the zeta potential generally varies with position along the surface (unlike in our previous work, where it was assumed spatially uniform), and so here we use the area-averaged value.  We normalize the electric fields by $i / \sigma$, where $i$ is the area-averaged current density out of the anode and $\sigma$ is the conductivity at the lowest salt concentration (yielding a conductivity of 8.8 $\mu$S/cm).  Since it is measured roughly one Debye length away from the surface, $E_{50}$ is an estimate of the electric field driving the electroviscous flow in the EDL and, by extension, rod motion.  However, the electric field varies significantly with position throughout the simulation domain, and is significantly more complicated than the relatively uniform electric field applied for conventional electrophoresis.  To get an idea of how the electric field as a whole acts on the particle, we focus our attention on the variation of $E^*$ with respect to conductivity.  Power-law curve fitting indicates that $E^*$ is roughly inversely proportional to conductivity, scaling as $E^* \propto \sigma^{-0.988}$.

\begin{figure}
\centering
\input{fig6.tex}
\caption{Deviation of the total reaction rate $J$ from its maximum value $J_{max}$, where $J_{max} = 1.21 \times 10^{-16}$~mol/s and is achieved at the minimum conductivity of 8.8 $\mu$S/cm.  Ionic strength was varied using KCl.  This figure shows that the addition of electrolyte decreases the overall reaction rate by approximately 20 \% at the highest ionic strength.  The ($\blacksquare$) symbols indicate simulation data, while the solid line is a power-law fit to the data.  The addition of electrolyte alters the proton concentration in the diffuse layer, altering the reaction rate on the cathode and leading to a net decrease in the reaction rate on both ends.}
\label{fig6}
\end{figure}
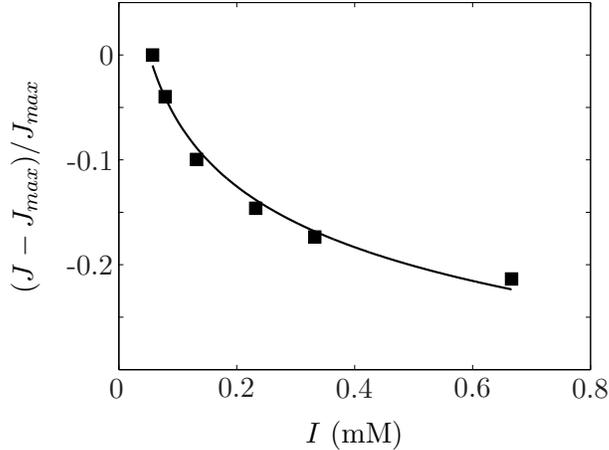

Figure \ref{fig6} shows that the normalized total reaction rate on the rod surface decreases as a function of solution ionic strength.  Here, $J_{max}$ is the total reaction rate at the minimum ionic strength and $J$ is the integrated flux over the surface of one electrode, defined in (\ref{currcons}), and depends on ionic strength.  The figure illustrates a 20 \% decrease in reaction rate from the minimum to the maximum conductivity, suggesting that the overall current through the fluid and the rod (due to all possible transport mechanisms) decreases with increased conductivity.  On a close examination of equation (\ref{fluxbc}), we see that the variables influencing the reaction rates are the concentration of peroxide, concentration of protons (in the case of Au), and the Stern-layer voltage.  Since the peroxide concentration is kept constant in all simulations, the variation in reaction rate with ionic strength is attributed to the Stern voltage and the concentration of protons near the Au surface.

\begin{figure}
\centering
\input{fig7.tex}
\caption{Magnitude of the Stern voltage (normalized by rod potential, $\Phi_{rod}$) evaluated at the midpoint of the rod as a function of the ratio of Stern layer thickness to Debye thickness.  The ratio $\lambda_S / \lambda_D$ is varied by keeping $\lambda_S$ constant at 2~{\AA} and varying $\lambda_D$ by adding KCl.  This plot supports the scaling prediction that the magnitude of the Stern voltage scales linearly with the ratio of Stern-layer thickness to Debye thickness.  ($\blacksquare$) symbols indicate simulation data, and the solid line indicates the linear scaling predicted by (\ref{scaling}).}
\label{fig7}
\end{figure}
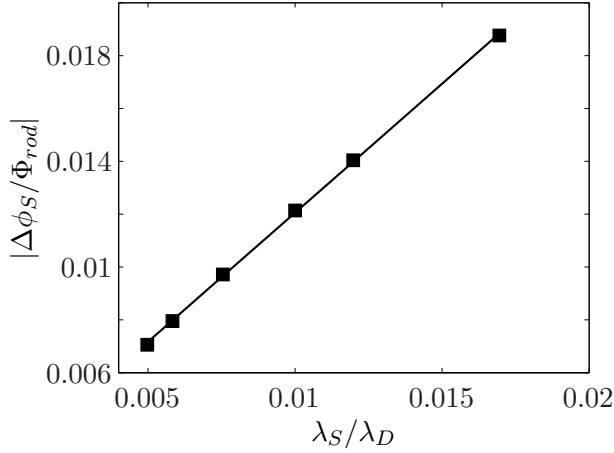

Figure \ref{fig7} shows the Stern voltage measured at the midpoint of the rod normalized by rod potential $\Phi_{rod}$ as a function of the ratio $\lambda_S / \lambda_D$, showing a linear relationship.  Considering that $\lambda_S$ is a constant while $\lambda_D$ depends inversely on the square root of ionic strength, the abscissa in Fig. \ref{fig7} is proportional to the square root of ionic strength.  Figure \ref{fig7} suggests that the Stern voltage becomes more important in relation to the rod potential as salt concentration is increased.

To understand why the Stern voltage scales with salt concentration, we refer to our previous scaling result for the Stern voltage as a function of the Stern-layer thickness, Debye thickness, and average zeta potential of the particle, derived from the Debye-H\"{u}ckel solution for a flat plate with zeta potential $\zeta$,\cite{moran_electrokinetic_2011}
\begin{equation}
\left| \Delta \phi_S \right|  \propto \frac{\lambda_S}{\lambda_D} \left| \zeta \right|.
\label{DphiSscale_old}
\end{equation}
For the bimetallic rods, area-averaged zeta potential scales as the internal potential $\Phi_{rod}$ and we expect
\begin{equation}
\left| \frac{\Delta \phi_S}{\Phi_{rod}} \right| \propto \frac{\lambda_S}{\lambda_D}.
\label{DphiSscale}
\end{equation}
The agreement between Fig. \ref{fig7} and (\ref{DphiSscale}) is robust, considering that (\ref{DphiSscale}) assumes that the rod potential is less than or comparable to $RT/F$, an assumption which is not rigorously justified at the values of $\Phi_{rod}$ calculated for the rods, which can be close to three times the thermal voltage (see Fig. \ref{fig4}).

\begin{figure}
\centering
\input{fig8.tex}
\caption{Variation of Stern voltage magnitude (normalized again by the thermal voltage, $RT/F$) with position along the rod at the indicated conductivities.  Conductivity increases from the bottom to the top curve, and is varied by adding KCl.  The left half of the plot ($z < 0$) corresponds to the Au segment of the rod, the right half ($z > 0$) to the Pt segment.  Stern voltage is controlled by the Stern-layer thickness and electric field normal to the rod surface.  As conductivity is increased, the diffuse layer shrinks, steepening the potential gradient at the surface and increasing the Stern voltage.}
\label{fig8}
\end{figure}
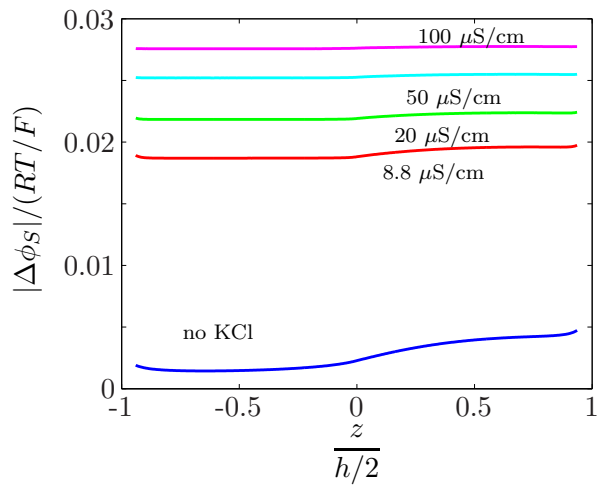

Figure \ref{fig8} shows the Stern voltage magnitude $|\Delta \phi_S|$ (defined in (\ref{potentialBC}) as a function of position along the rod at the indicated conductivities.  The Stern voltage clearly increases in magnitude with conductivity as predicted.  The variations in Stern voltage in Fig. \ref{fig8} arise from variations of the normal electric field along the surface (see Eq. (\ref{potentialBC})).  The radial electric field varies with $z$ due to the asymmetric production and consumption of protons along the rod.  On the anode side ($z > 0$), protons are generated, making the electric potential in the fluid immediately adjacent to the rod marginally more positive compared to the fluid adjacent to the cathode side ($z < 0$), where protons are consumed.  As a result, the potential gradient across the Stern layer is slightly larger near the anode, which increases the normal electric field and leads to a larger magnitude of Stern voltage.  Due to the elevated electrolyte concentration, the Stern voltage is significantly greater in magnitude than in our previous work (where it was near $10^{-6}$~V), in which we do not consider added electrolytes.\cite{moran_electrokinetic_2011}

This analysis shows that as more electrolyte is added to the solution, the shrinking diffuse layer causes stronger radial electric fields at the particle surface, leading to a larger Stern voltage.  The exponential terms in the Butler-Volmer rate equations therefor deviate further from unity as electrolyte concentration is increased.  However, even at the highest salt concentration, the exponential terms are bounded by 0.97 and 1.03, and so exert relatively little influence on the reaction kinetics.

\begin{figure}
\centering
\input{fig9a.tex}
\input{fig9b.tex}
\caption{(a) Non-dimensionalized perturbation of proton concentration about the bulk value, $c_{+,\infty} = 9 \times 10^{-7}$~mol/L, as a function of $z$ along the gold (cathode) surface of the rod for two conductivities: 8.8 $\mu$S/cm (solid line) and 100 $\mu$S/cm (dashed line).  Conductivities were varied by adjusting KCl concentration.  The proton concentration along gold decreases slightly with increasing conductivity, lowering the reaction rate on the gold due to the quadratic dependence of the cathodic reaction rate on proton concentration.  (b) Ratio of the square of proton concentration at the high conductivity (100 $\mu$S/cm) to the square of proton concentration at the low conductivity (8.8~$\mu$S/cm), as a function of position along the gold surface.  This plot demonstrates that the contribution of proton concentration in the high-salt case to the reaction rate is significantly lower than in the low-salt case, especially near the Pt/Au junction.  As a result, the reaction rate on the cathode decreases due to the local decrease in proton concentration.}
\label{fig9}
\end{figure}
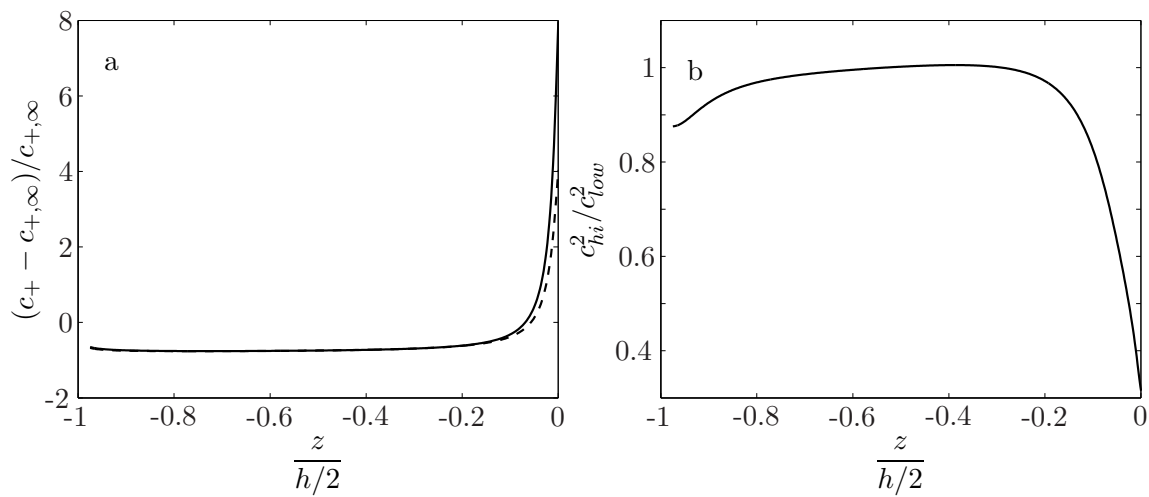

The proton concentration in the diffuse layer directly affects the reaction rate on the cathode, as shown in the kinetic expressions (\ref{fluxbc}).  Figure \ref{fig9} (a) shows the concentration of protons along the cathode surface as a function of position for two separate conductivities.  At the higher conductivity, the concentration is reduced, especially near the junction ($z = 0$).  As electrolyte is added, more of the counterions making up the electrolyte (in this case, potassium cations, K$^+$) move into the diffuse layer to help screen the negative surface charge.  Since the electrolyte is two orders of magnitude more concentrated than the protons, the K$^+$ ions alleviate the need for protons to screen the surface charge.  As a result, the proton concentration is diminished.

Although the reduction in proton concentration appears relatively insignificant, Fig. \ref{fig9} (b) shows that it can still have a noticeable effect on the reaction kinetics.  In Fig. \ref{fig9} (b) we show the ratio of the square of proton concentration for the high and low-conductivity cases.  Since the cathodic reaction rate depends quadratically on proton concentration, this figure effectively shows the reduction in reaction rate, as a function of position, due to the reduction in proton concentration.  The reduction is significant near the anode/cathode junction (close to $z = 0$), showing that the decrease in proton concentration with bulk electrolyte concentration exerts a non-negligible influence on the total reaction rate, and partially explains why the total reaction rate decreases with ionic strength (Fig. \ref{fig6}).

\subsection{Dimensionless Parameters}
To gain an understanding of the relative dominance of different transport mechanisms in the system, we calculate the values of the Damk\"{o}hler and Rayleigh numbers and the parameter $\beta$ from their respective formulas using appropriate values of the system variables.  In addition, we estimate the parameters using values we extract from the simulations.  Specifically, we calculate approximations to these three dimensionless parameters by numerically integrating the relevant fluxes throughout the system geometry and computing the relevant ratios.  Table \ref{salttab2} shows the results of these calculations.

\input{salttab2.tex}

The total reaction-driven flux of protons is given $J$, defined in (\ref{currcons}).  The total diffusive, convective and electromigration fluxes of protons through the fluid are estimated by integrating the $z$-components of the appropriate local fluxes over the two-dimensional annular disk surrounding the middle of the rod and extending from the rod surface to the boundary of the simulation domain.  The parameters are then numerically estimated by computing the relevant flux ratios: convective to diffusive flux ($Ra_e$), electromigration to diffusive ($\beta$), and total reaction flux to diffusive flux ($Da$).

Although the numerical values of the dimensionless parameters differ significantly between the analytical to the numerical versions, the general trends are the same in both cases.  The Rayleigh number is $O(10^{-4})$ in the analytical case and $O(10^{-1})$ in the numerical case, indicating that electroconvection is dominated by diffusion in transporting mass.  The values of $\beta$ are no larger than 0.01 in either case, showing that electromigration is relatively unimportant compared to diffusion.  The Damk\"{o}hler numbers are on similar orders of magnitude for both cases, suggesting that the majority of the charged species injected into the solution by the reactions are transported by diffusion.

In all cases, the dimensionless parameters become smaller as salt concentration is increased.  In the case of the Rayleigh number, this decrease reflects the decreased swimming speed of the rod, reducing the convective flux of all species.  The reduction in $\beta$ is due to the reduction in the tangential electric field magnitude, which drives electromigration flux.  Finally, the Damk\"{o}hler numbers decrease with salt concentration, indicating that the reaction rate decreases slightly as salt is added.  Overall, we conclude from this analysis that the bulk of transport in the system is due to diffusion (to a greater extent than in the case without salt\cite{moran_electrokinetic_2011}), and electroconvection is relatively unimportant in transporting species.

\subsection{Physical Reasons for Speed Decrease}
Figure \ref{fig2} shows that the rod velocity decreases roughly by a factor of 20 when the conductivity is increased from 8.8 to 100 $\mu$S/cm.  This velocity decrease is due to several factors, each of which cause a partial reduction in the swimming speed.  In descending order of importance, these factors are (i) the decrease in magnitude of the propulsive electric field, (ii) the decrease in magnitude of the area-averaged zeta potential, and (iii) the decrease in overall reaction rate.

We calculated the characteristic electric field $E^*$ of the system as the electric field that would need to be externally applied to drive conventional electrophoresis of the rod having the same surface charge at a speed equal to the measured swimming speed.  As conductivity is increased from 8.8 to 100 $\mu$S/cm, $E^*$ decreases from 415 to 38 V/m, a decrease of roughly 90 \%, i.e. $E^* (\sigma = 8.8) / E^* (\sigma = 100)$ = 10.92.  Since swimming speed is directly proportional to $E^*$, the reduction in $E^*$ with conductivity reduces the swimming speed of the rod by roughly an order of magnitude.

In all cases, the electrolyte added to the solution is at a significantly higher bulk concentration than protons and bicarbonate ions.  While the bulk concentration of protons and bicarbonate ions was kept fixed at 0.9 $\mu$mol/L, the electrolyte concentration was varied from 56.4 to 820 $\mu$mol/L.  The counterions in the electrolyte are attracted to the rod to help screen the surface charge and result in a net decrease in proton concentration in the diffuse layer, as shown in Fig. \ref{fig9} (a).  The reduced proton concentration causes a reduction in the reaction rate because the reaction rate on the gold is dependent on the square of the proton concentration.  When conductivity is increased from its minimum to its maximum value, the total reaction rates on the anode and cathode decrease by roughly 20 \%, such that $J (\sigma = 8.8) / J (\sigma = 100)$ = 1.27.  Due to the proportionality between the speed and surface activity identified by Golestanian, Liverpool, and Ajdari,\cite{golestanian_designing_2007} and also proposed and validated in the scaling analysis of our previous work,\cite{moran_locomotion_2010,moran_electrokinetic_2011} we conclude that the reduction in reaction rate causes a decrease in swimming speed of roughly 20 \% from the minimum to the maximum conductivity.

Figure \ref{fig4} shows that the average zeta potential decreases with the addition of salt.  This change in $\zeta$, which is closely linked to the rod potential, is driven by the decrease in reaction rates and the conservation of current requirement.  The reaction rate on the cathode is reduced as salt is added, due to the limited availability of protons (see Fig. \ref{fig9}) to participate in the peroxide reduction reaction.  Since current must be conserved, the rod potential self-adjusts such that the rate of peroxide oxidation on the anode decreases.  The average zeta potential decreases by roughly 45 \%, from $-$67.9 to $-$37.1 mV.  $\bar{\zeta} (\sigma = 8.8) / \bar{\zeta} (\sigma = 100) = 1.83$.  The change in zeta potential should result in a reduction in the swimming speed by roughly a factor of 2.

Since speed is linearly related to $E^*$, zeta potential, and reaction rate, we can multiply the reduction factors together to obtain an estimate of the total velocity reduction factor due to all three effects.  The result is 25.38, which is in reasonably good agreement with the observed speed reduction factor of $U (\sigma = 8.8) / U (\sigma = 100) = 19.86$.

In summary, the conductivity-induced speed decrease originates from the decrease in the characteristic electric field (due to Ohm's law), and to a lesser degree, the reduction of the cathode reaction rate due to exclusion of protons in the diffuse layer by the added nonreactive salt. 

%% file: fig1a.tex
\begin{psfrags}%
\psfragscanon%
\psfrag{a}[cc][c]{(a)}%
\psfrag{2}[cc][c]{2}%
\psfrag{1}[cc][c]{1}%
\psfrag{f}[cc][c]{-2}%
\psfrag{g}[cc][c]{-1}%
\psfrag{x}[cc][c]{$r/h$}%
\psfrag{y}[cc][c]{$z/h$}%
\psfrag{0}[cc][c]{0}%
\includegraphics[clip,width=3.0in]{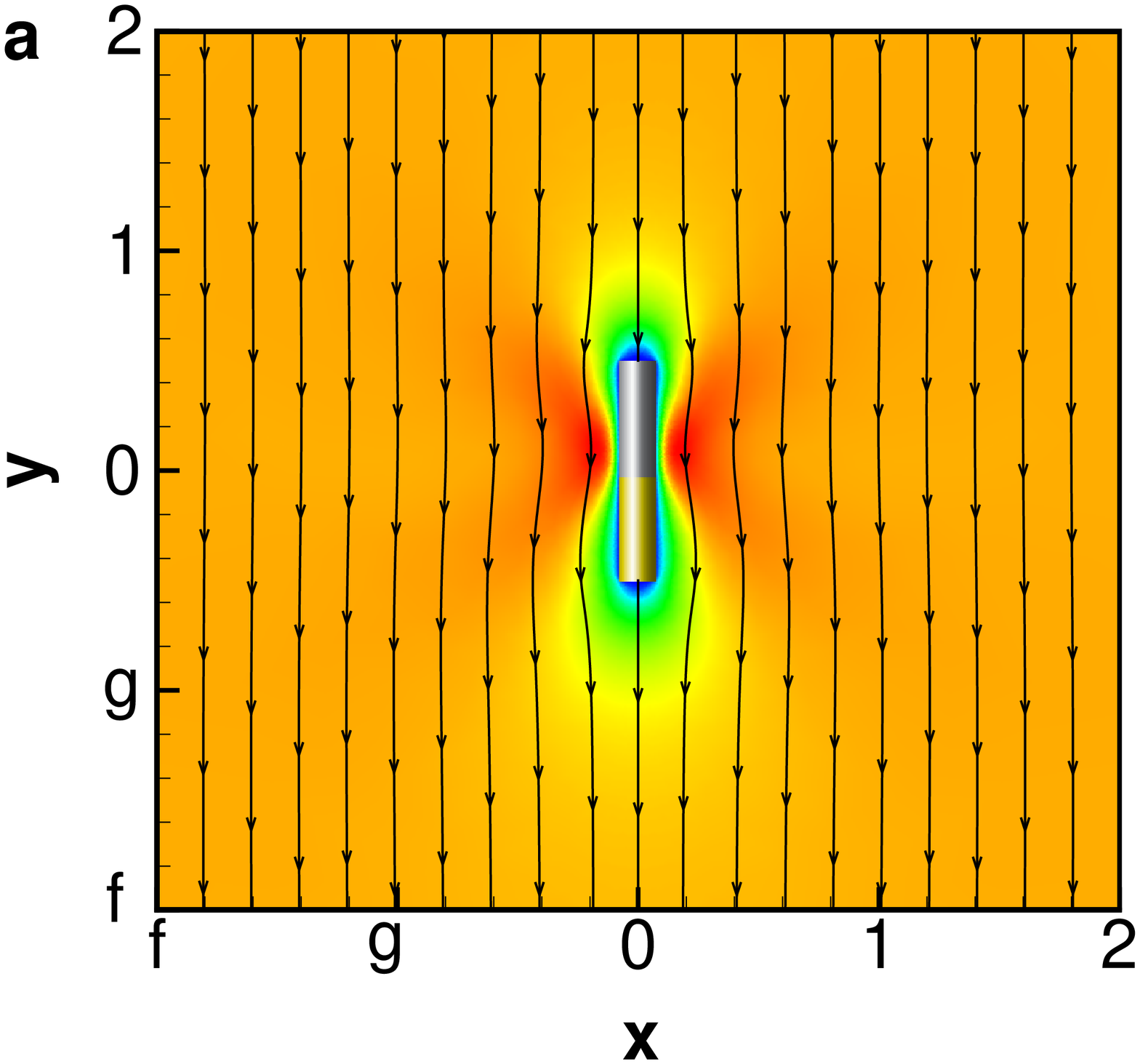}%
\end{psfrags}%

%% file: fig1b.tex
\begin{psfrags}%
\psfragscanon%
\psfrag{b}[cc][c]{(b)}%
\psfrag{2}[cc][c]{2}%
\psfrag{1}[cc][c]{1}%
\psfrag{x}[cc][c]{$r/h$}%
\psfrag{y}[cc][c]{$z/h$}%
\psfrag{0}[cc][c]{0}%
\psfrag{f}[cc][c]{$-2$}%
\psfrag{h}[cc][c]{$-1$}%
\includegraphics[clip,width=3.0in]{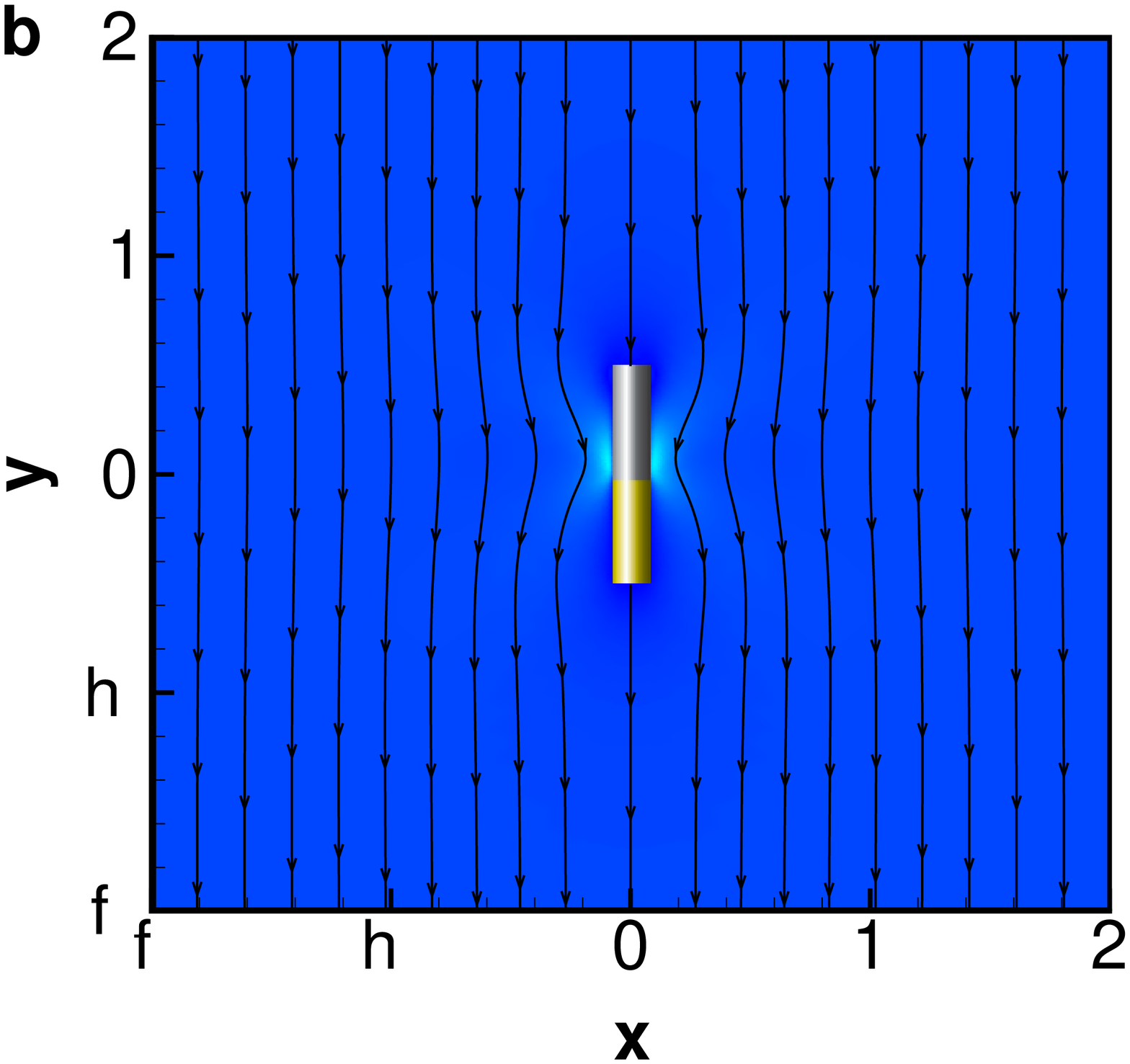}%
\end{psfrags} %

%% file: fig2a.tex
%
%
\begin{psfrags}%
\psfragscanon%
%
\psfrag{s03}[t][t]{\color[rgb]{0,0,0}\setlength{\tabcolsep}{0pt}\begin{tabular}{c}$\sigma$ ($\mu$S/cm)\end{tabular}}%
\psfrag{s04}[b][b]{\color[rgb]{0,0,0}\setlength{\tabcolsep}{0pt}\begin{tabular}{c}$U$ ($\mu$m/s)\end{tabular}}%
\psfrag{s05}[l][l]{\color[rgb]{0,0,0}\setlength{\tabcolsep}{0pt}\begin{tabular}{l}a\end{tabular}}%
%
\psfrag{x01}[t][t]{0}%
\psfrag{x02}[t][t]{20}%
\psfrag{x03}[t][t]{40}%
\psfrag{x04}[t][t]{60}%
\psfrag{x05}[t][t]{80}%
\psfrag{x06}[t][t]{100}%
%
\psfrag{v01}[r][r]{0}%
\psfrag{v02}[r][r]{5}%
\psfrag{v03}[r][r]{10}%
\psfrag{v04}[r][r]{15}%
\psfrag{v05}[r][r]{20}%
\psfrag{v06}[r][r]{25}%
\psfrag{v07}[r][r]{30}%
%
\includegraphics[clip,width=3.1in]{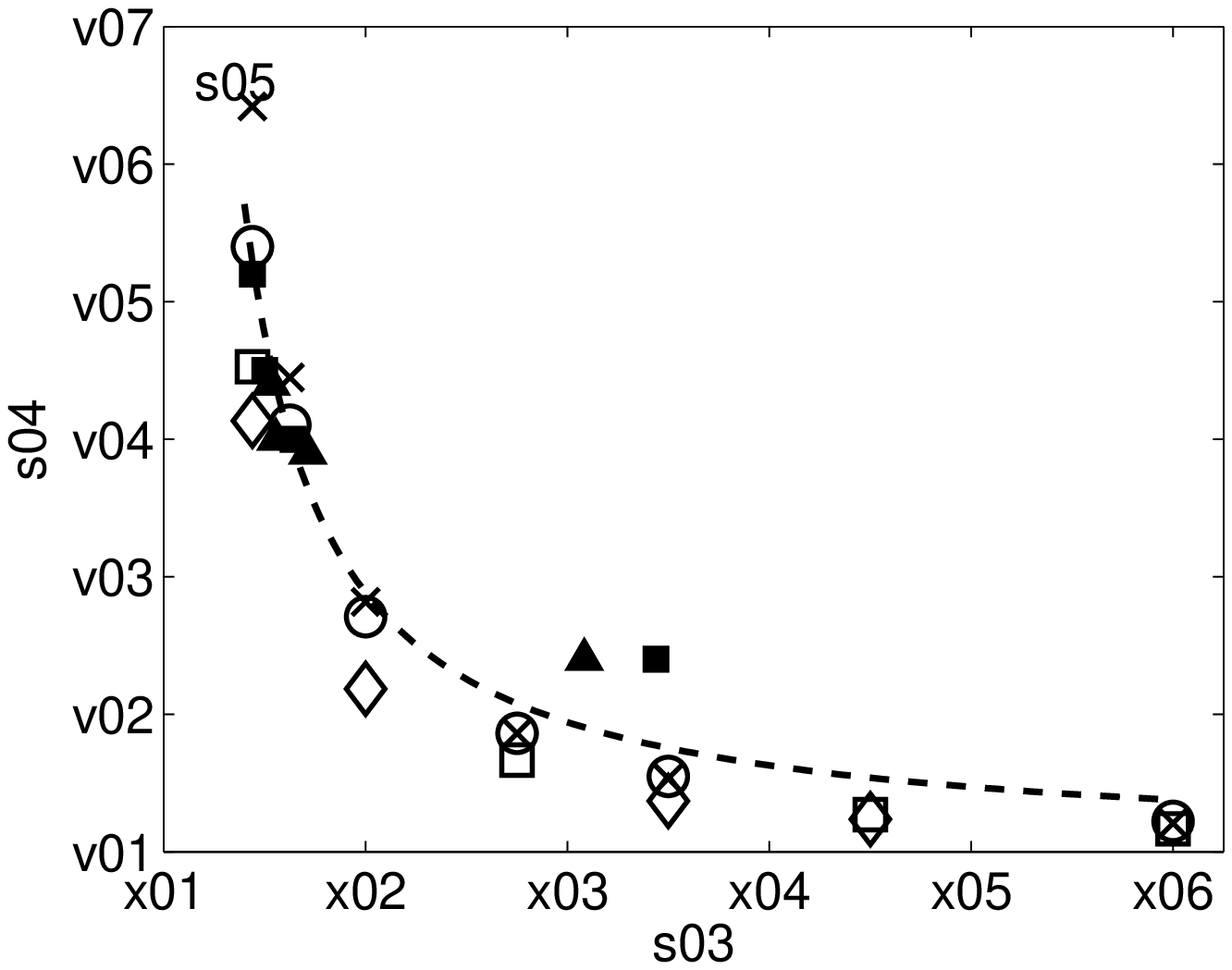}
\end{psfrags}%
%

%% file: fig2b.tex
%
%
\begin{psfrags}%
\psfragscanon%
%
\psfrag{s04}[l][l]{\color[rgb]{0,0,0}\setlength{\tabcolsep}{0pt}\begin{tabular}{l}b\end{tabular}}%
\psfrag{s05}[t][t]{\color[rgb]{0,0,0}\setlength{\tabcolsep}{0pt}\begin{tabular}{c}$I$ (mM)\end{tabular}}%
%
\psfrag{x01}[t][t]{0}%
\psfrag{x02}[t][t]{0.2}%
\psfrag{x03}[t][t]{0.4}%
\psfrag{x04}[t][t]{0.6}%
\psfrag{x05}[t][t]{0.8}%
\psfrag{x06}[t][t]{1}%
%
\psfrag{v01}[r][r]{0}%
\psfrag{v02}[r][r]{5}%
\psfrag{v03}[r][r]{10}%
\psfrag{v04}[r][r]{15}%
\psfrag{v05}[r][r]{20}%
\psfrag{v06}[r][r]{25}%
\psfrag{v07}[r][r]{30}%
%
\includegraphics[clip,width=3.01in]{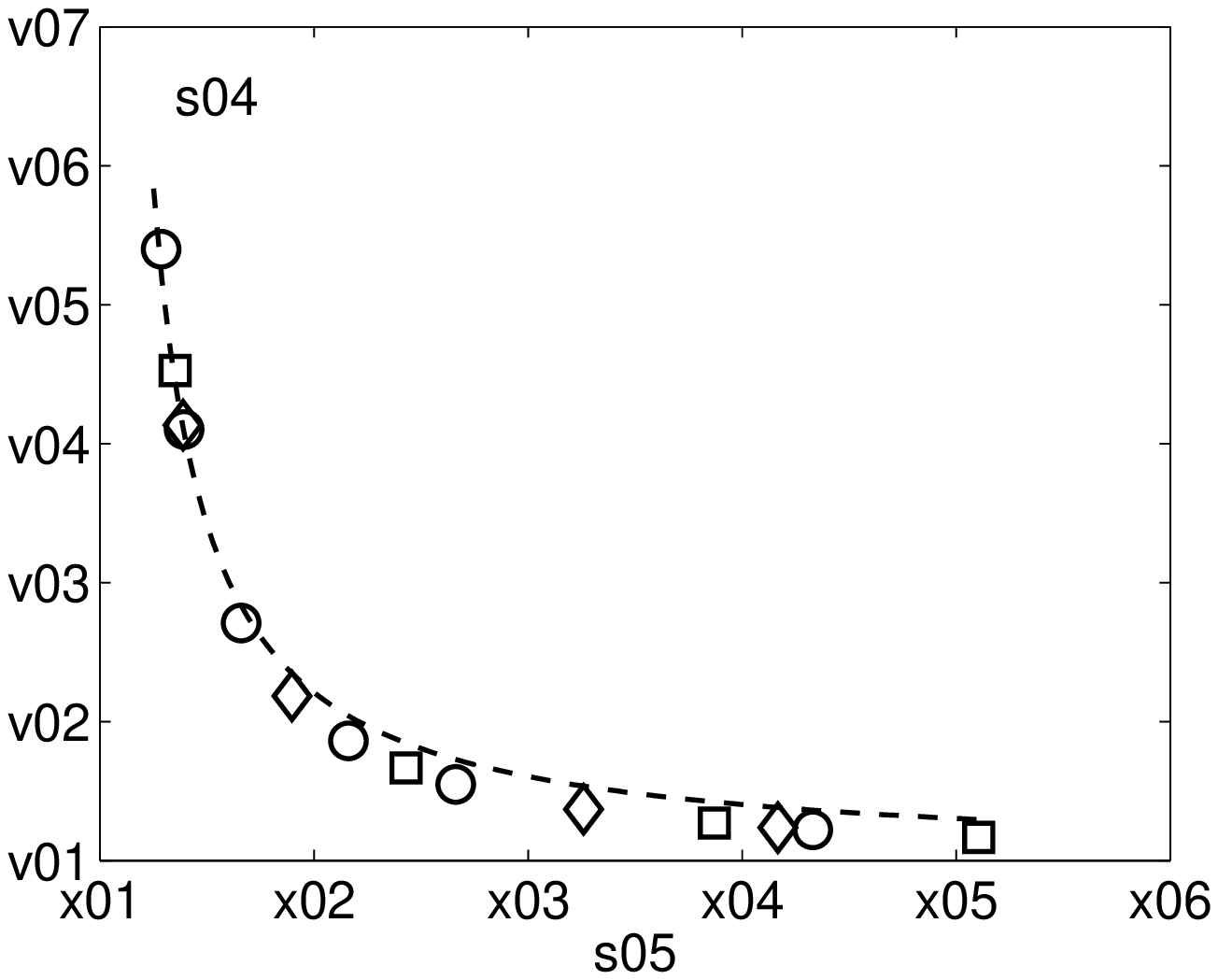}
\end{psfrags}%
%

%% file: fig3.tex
%
%
\begin{psfrags}%
\psfragscanon%
%
\psfrag{s03}[t][t]{\color[rgb]{0,0,0}\setlength{\tabcolsep}{0pt}\begin{tabular}{c}$\lambda_D$ (nm)\end{tabular}}%
\psfrag{s04}[b][b]{\color[rgb]{0,0,0}\setlength{\tabcolsep}{0pt}\begin{tabular}{c}$U$ ($\mu$m/s)\end{tabular}}%
%
\psfrag{x01}[t][]{10}%
\psfrag{x02}[t][]{15}%
\psfrag{x03}[t][]{20}%
\psfrag{x04}[t][]{25}%
\psfrag{x05}[t][]{30}%
\psfrag{x06}[t][]{35}%
\psfrag{x07}[t][]{40}%
%
\psfrag{v01}[r][r]{0}%
\psfrag{v02}[r][r]{5}%
\psfrag{v03}[r][r]{10}%
\psfrag{v04}[r][r]{15}%
\psfrag{v05}[r][r]{20}%
\psfrag{v06}[r][r]{25}%
%
\includegraphics[clip,width=3in]{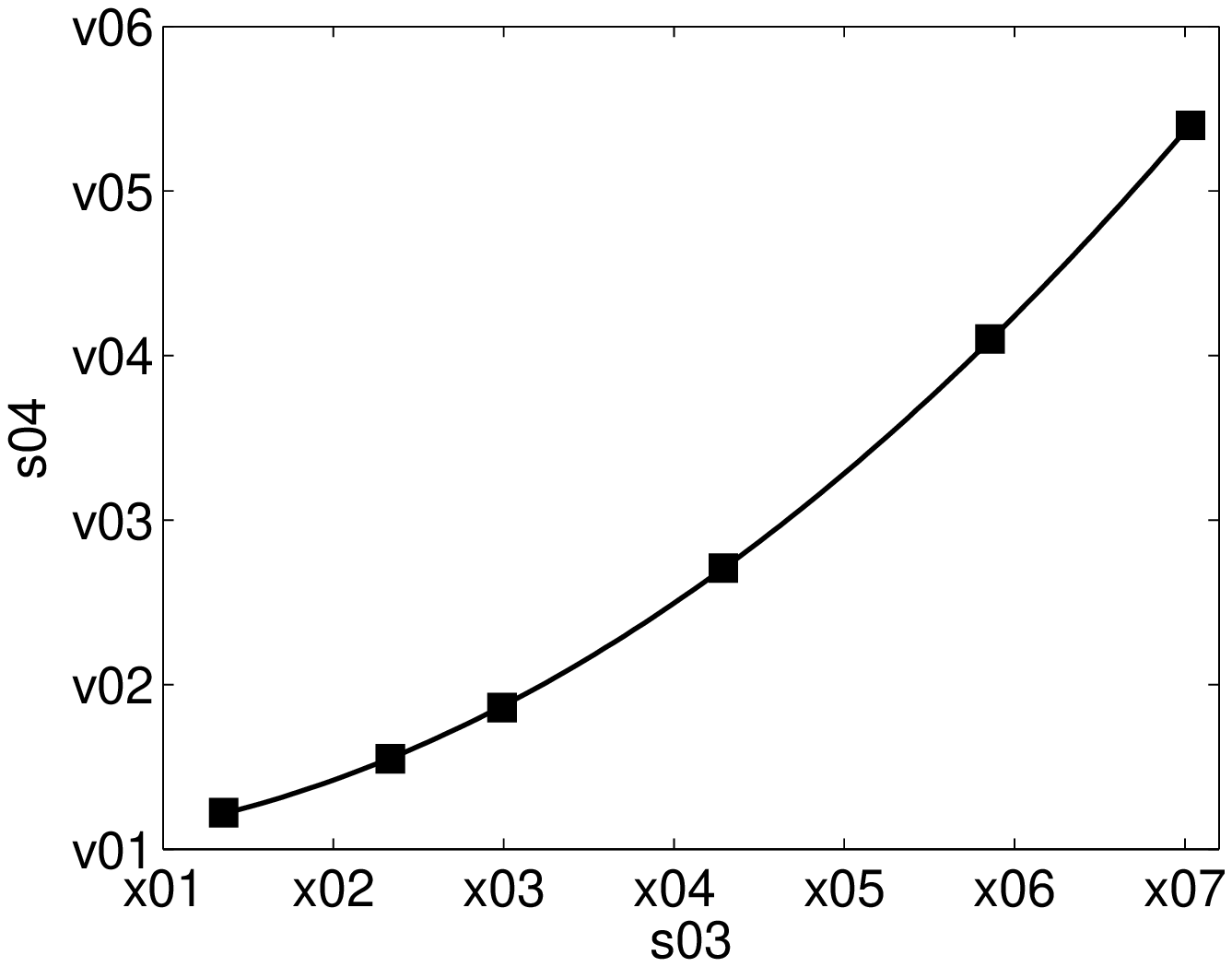}
\end{psfrags}%
%

%% file: fig4a.tex
%
%
\begin{psfrags}%
\psfragscanon%
%
\psfrag{s03}[t][t]{\color[rgb]{0,0,0}\setlength{\tabcolsep}{0pt}\begin{tabular}{c}$\sigma$ ($\mu$S/cm)\end{tabular}}%
\psfrag{s04}[b][b]{\color[rgb]{0,0,0}\setlength{\tabcolsep}{0pt}\begin{tabular}{c}$\zeta_{avg} / (RT/F)$\end{tabular}}%
\psfrag{s05}[l][l]{\color[rgb]{0,0,0}\setlength{\tabcolsep}{0pt}\begin{tabular}{l}a\end{tabular}}%
%
\psfrag{x01}[t][t]{0}%
\psfrag{x02}[t][t]{20}%
\psfrag{x03}[t][t]{40}%
\psfrag{x04}[t][t]{60}%
\psfrag{x05}[t][t]{80}%
\psfrag{x06}[t][t]{100}%
%
\psfrag{v01}[r][r]{-3}%
\psfrag{v02}[r][r]{-2.5}%
\psfrag{v03}[r][r]{-2}%
\psfrag{v04}[r][r]{-1.5}%
\psfrag{v05}[r][r]{-1}%
%
\includegraphics[clip,width=3.1in]{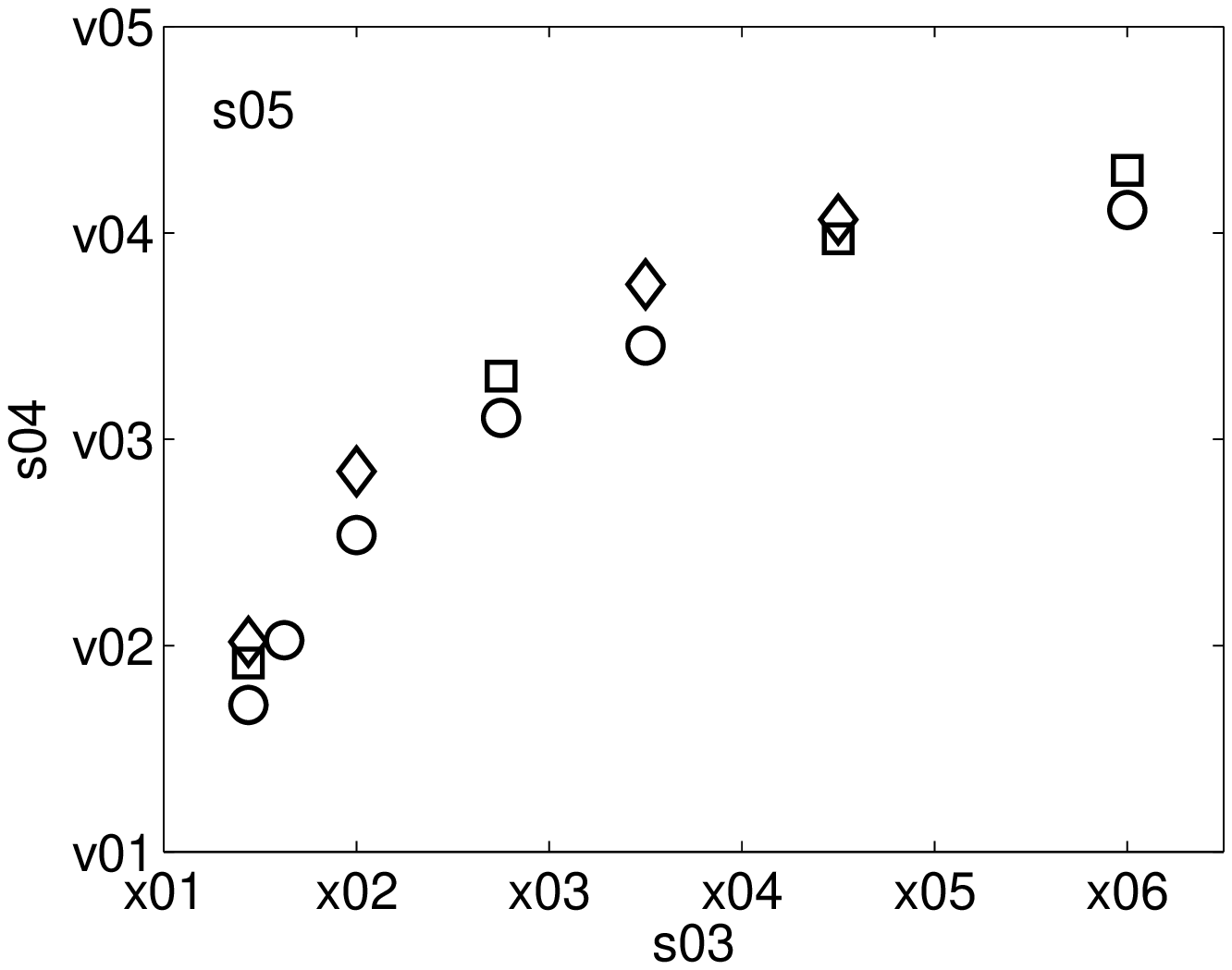}
\end{psfrags}%
%

%% file: fig4b.tex
%
%
\begin{psfrags}%
\psfragscanon%
%
\psfrag{s04}[t][t]{\color[rgb]{0,0,0}\setlength{\tabcolsep}{0pt}\begin{tabular}{c}$I$ (mM)\end{tabular}}%
\psfrag{s05}[l][l]{\color[rgb]{0,0,0}\setlength{\tabcolsep}{0pt}\begin{tabular}{l}b\end{tabular}}%
%
\psfrag{x01}[t][t]{0}%
\psfrag{x02}[t][t]{0.2}%
\psfrag{x03}[t][t]{0.4}%
\psfrag{x04}[t][t]{0.6}%
\psfrag{x05}[t][t]{0.8}%
\psfrag{x06}[t][t]{1}%
%
\psfrag{v01}[r][r]{-3}%
\psfrag{v02}[r][r]{-2.5}%
\psfrag{v03}[r][r]{-2}%
\psfrag{v04}[r][r]{-1.5}%
\psfrag{v05}[r][r]{-1}%
%
\includegraphics[clip,width=3.01in]{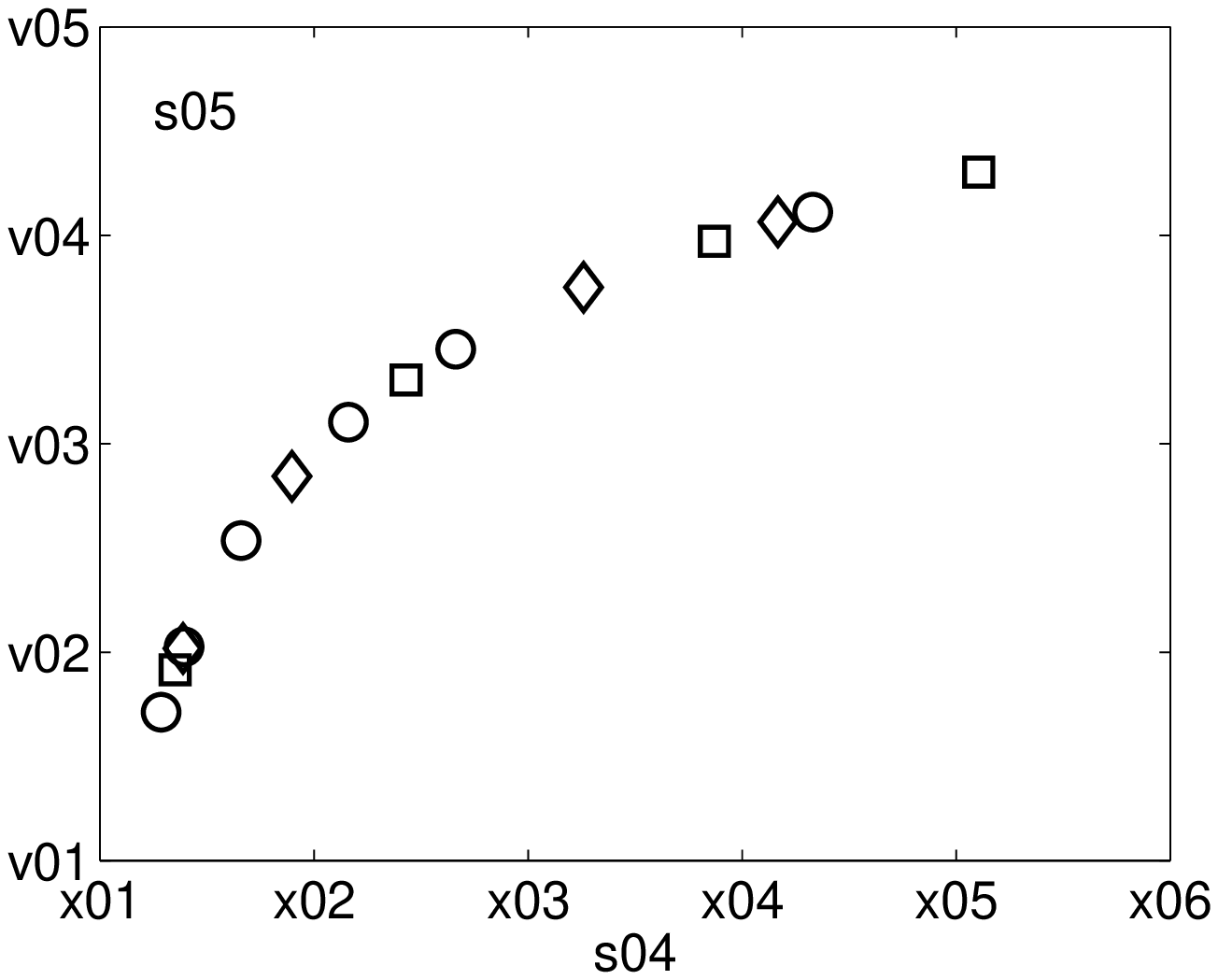}
\end{psfrags}%
%

%% file: fig5.tex
%
%
\begin{psfrags}%
\psfragscanon%
%
\psfrag{s03}[t][t]{\color[rgb]{0,0,0}\setlength{\tabcolsep}{0pt}\begin{tabular}{c}$\sigma$ ($\mu$S/cm)\end{tabular}}%
\psfrag{s04}[b][b]{\color[rgb]{0,0,0}\setlength{\tabcolsep}{0pt}\begin{tabular}{c}$E \sigma / i$\end{tabular}}%
%
\psfrag{x01}[t][t]{0}%
\psfrag{x02}[t][t]{20}%
\psfrag{x03}[t][t]{40}%
\psfrag{x04}[t][t]{60}%
\psfrag{x05}[t][t]{80}%
\psfrag{x06}[t][t]{100}%
%
\psfrag{v01}[r][r]{0}%
\psfrag{v02}[r][r]{0.05}%
\psfrag{v03}[r][r]{0.1}%
\psfrag{v04}[r][r]{0.15}%
\psfrag{v05}[r][r]{0.2}%
\psfrag{v06}[r][r]{0.25}%
%
\includegraphics[clip,width=3in]{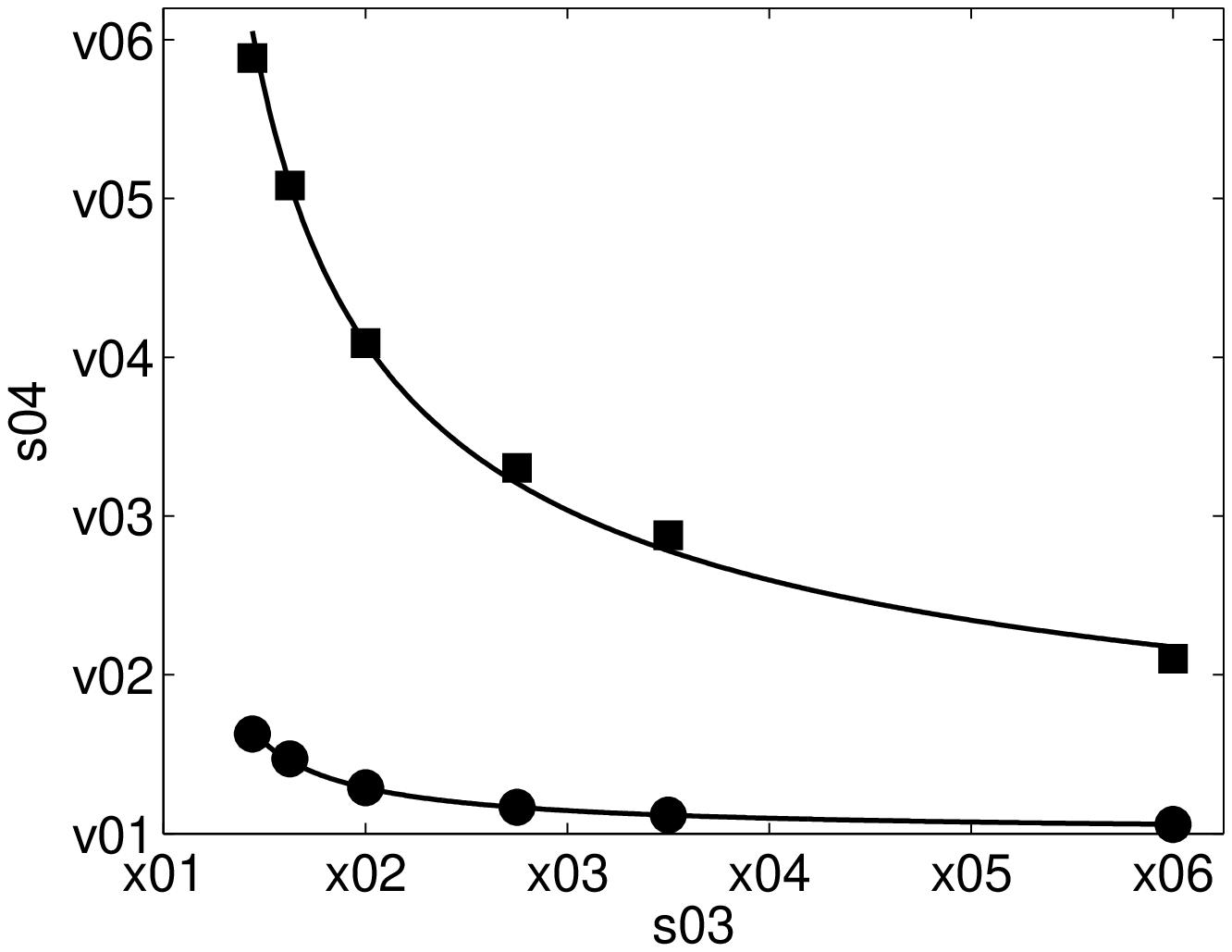}
\end{psfrags}%
%

%% file: fig6.tex
%
%
\begin{psfrags}%
\psfragscanon%
%
\psfrag{s03}[t][t]{\color[rgb]{0,0,0}\setlength{\tabcolsep}{0pt}\begin{tabular}{c}$I$ (mM)\end{tabular}}%
\psfrag{s04}[b][t]{\color[rgb]{0,0,0}\setlength{\tabcolsep}{0pt}\begin{tabular}{c}$(J - J_{max}) / J_{max}$\end{tabular}}%
%
\psfrag{x01}[t][t]{0}%
\psfrag{x02}[t][t]{0.2}%
\psfrag{x03}[t][t]{0.4}%
\psfrag{x04}[t][t]{0.6}%
\psfrag{x05}[t][t]{0.8}%
%
\psfrag{v01}[r][r]{}%
\psfrag{v02}[r][r]{-0.2}%
\psfrag{v03}[r][r]{}%
\psfrag{v04}[r][r]{-0.1}%
\psfrag{v05}[r][r]{}%
\psfrag{v06}[r][r]{0}%
\psfrag{v07}[r][r]{}%
%
\includegraphics[clip,width=3in]{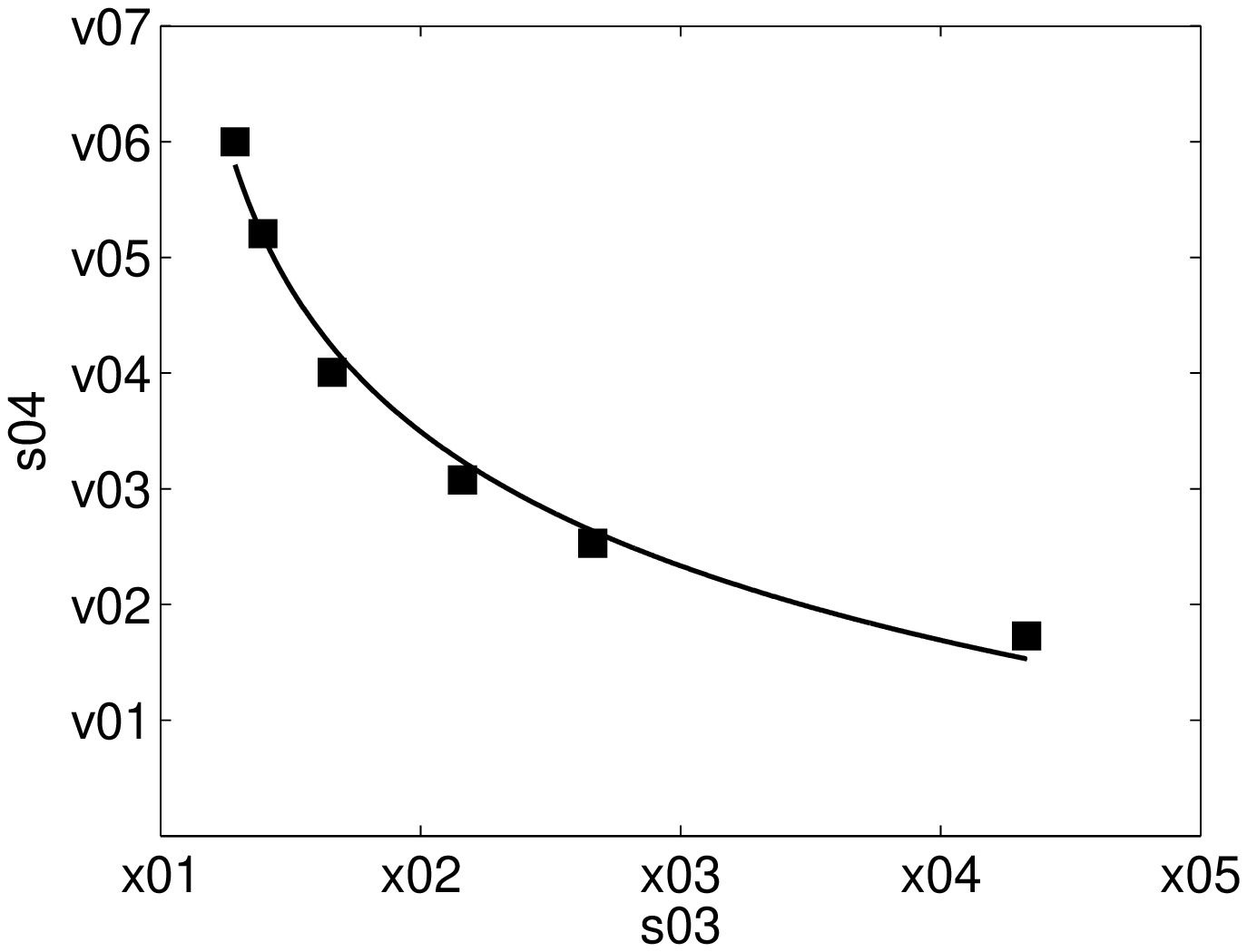}
\end{psfrags}%
%

%% file: fig7.tex

%
%
\begin{psfrags}%
\psfragscanon%
%
\psfrag{s03}[t][t]{\color[rgb]{0,0,0}\setlength{\tabcolsep}{0pt}\begin{tabular}{c}$\lambda_S / \lambda_D$\end{tabular}}%
\psfrag{s04}[b][t]{\color[rgb]{0,0,0}\setlength{\tabcolsep}{0pt}\begin{tabular}{c}$|\Delta\phi_S / \Phi_{rod}|$ \end{tabular}}%
%
\psfrag{x01}[t][]{0.005}%
\psfrag{x02}[t][]{0.01}%
\psfrag{x03}[t][]{0.015}%
\psfrag{x04}[t][]{0.02}%
%
\psfrag{v01}[r][r]{0.006}%
\psfrag{v02}[r][r]{}%
\psfrag{v03}[r][r]{0.01}%
\psfrag{v04}[r][r]{}%
\psfrag{v05}[r][r]{0.014}%
\psfrag{v06}[r][r]{}%
\psfrag{v07}[r][r]{0.018}%
\psfrag{v08}[r][r]{}%
%
\includegraphics[clip,width=3in]{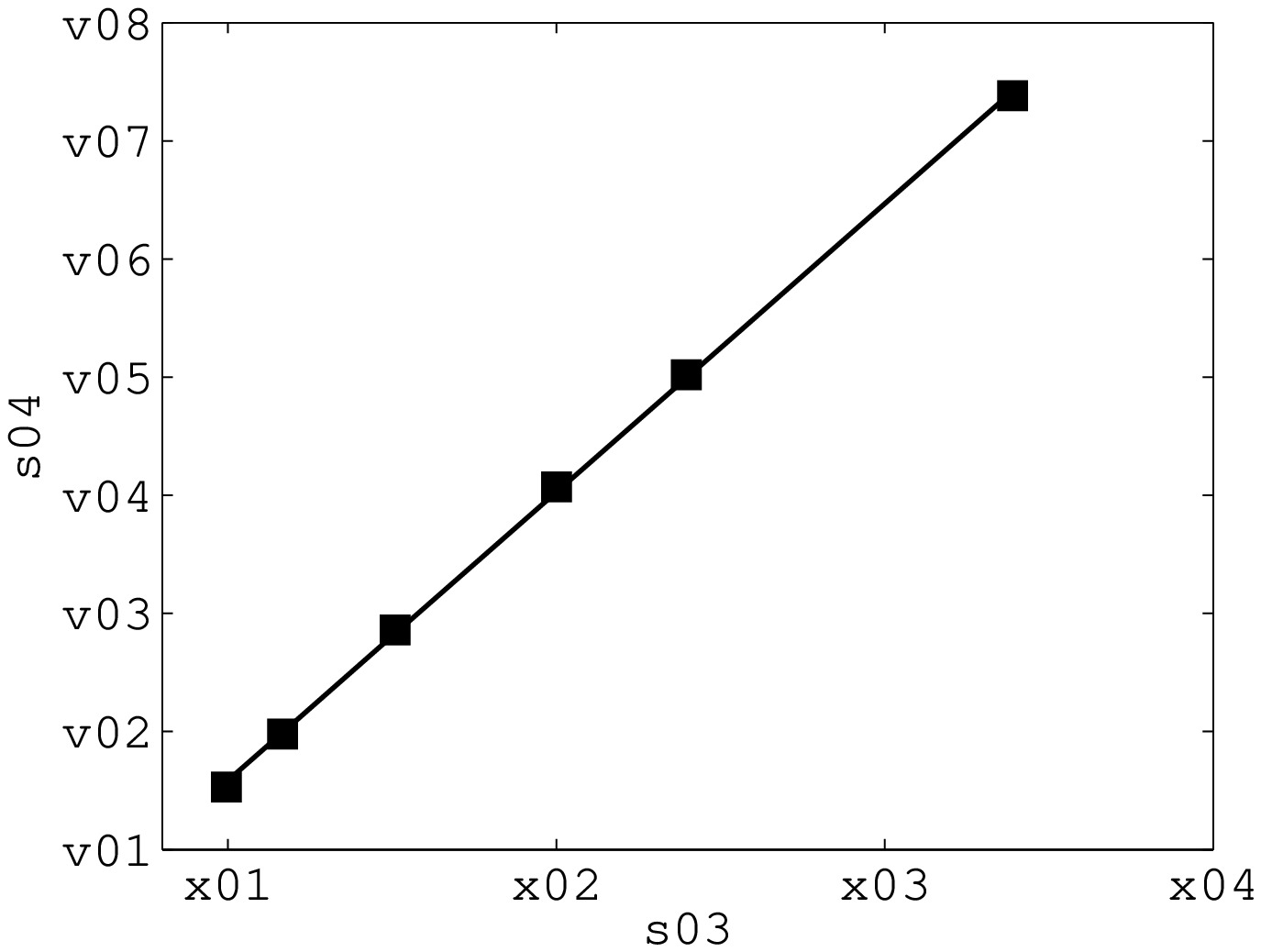}
\end{psfrags}%
%

%% file: fig8.tex

%
%
\begin{psfrags}%
\psfragscanon%
%
\psfrag{s03}[l][l]{\color[rgb]{0,0,0}\setlength{\tabcolsep}{0pt}\begin{tabular}{l}{\scriptsize no KCl}\end{tabular}}%
\psfrag{s04}[l][l]{\color[rgb]{0,0,0}\setlength{\tabcolsep}{0pt}\begin{tabular}{l}{\scriptsize 8.8 $\mu$S/cm}\end{tabular}}%
\psfrag{s05}[l][l]{\color[rgb]{0,0,0}\setlength{\tabcolsep}{0pt}\begin{tabular}{l}{\scriptsize 20 $\mu$S/cm}\end{tabular}}%
\psfrag{s06}[l][l]{\color[rgb]{0,0,0}\setlength{\tabcolsep}{0pt}\begin{tabular}{l}{\scriptsize 50 $\mu$S/cm}\end{tabular}}%
\psfrag{s07}[l][l]{\color[rgb]{0,0,0}\setlength{\tabcolsep}{0pt}\begin{tabular}{l}{\scriptsize 100 $\mu$S/cm}\end{tabular}}%
\psfrag{s08}[t][t]{\color[rgb]{0,0,0}\setlength{\tabcolsep}{0pt}\begin{tabular}{c}$\displaystyle \frac{z}{h/2}$\end{tabular}}%
\psfrag{s09}[b][t]{\color[rgb]{0,0,0}\setlength{\tabcolsep}{0pt}\begin{tabular}{c}{\ $|\Delta\phi_S| / (RT/F)$}\end{tabular}}%
%
\psfrag{x01}[t][t]{-1}%
\psfrag{x02}[t][t]{-0.5}%
\psfrag{x03}[t][t]{0}%
\psfrag{x04}[t][t]{0.5}%
\psfrag{x05}[t][t]{1}%
%
\psfrag{v01}[r][r]{0}%
\psfrag{v02}[r][r]{}%
\psfrag{v03}[r][r]{0.01}%
\psfrag{v04}[r][r]{}%
\psfrag{v05}[r][r]{0.02}%
\psfrag{v06}[r][r]{}%
\psfrag{v07}[r][r]{0.03}%
%
\includegraphics[clip,width=3in]{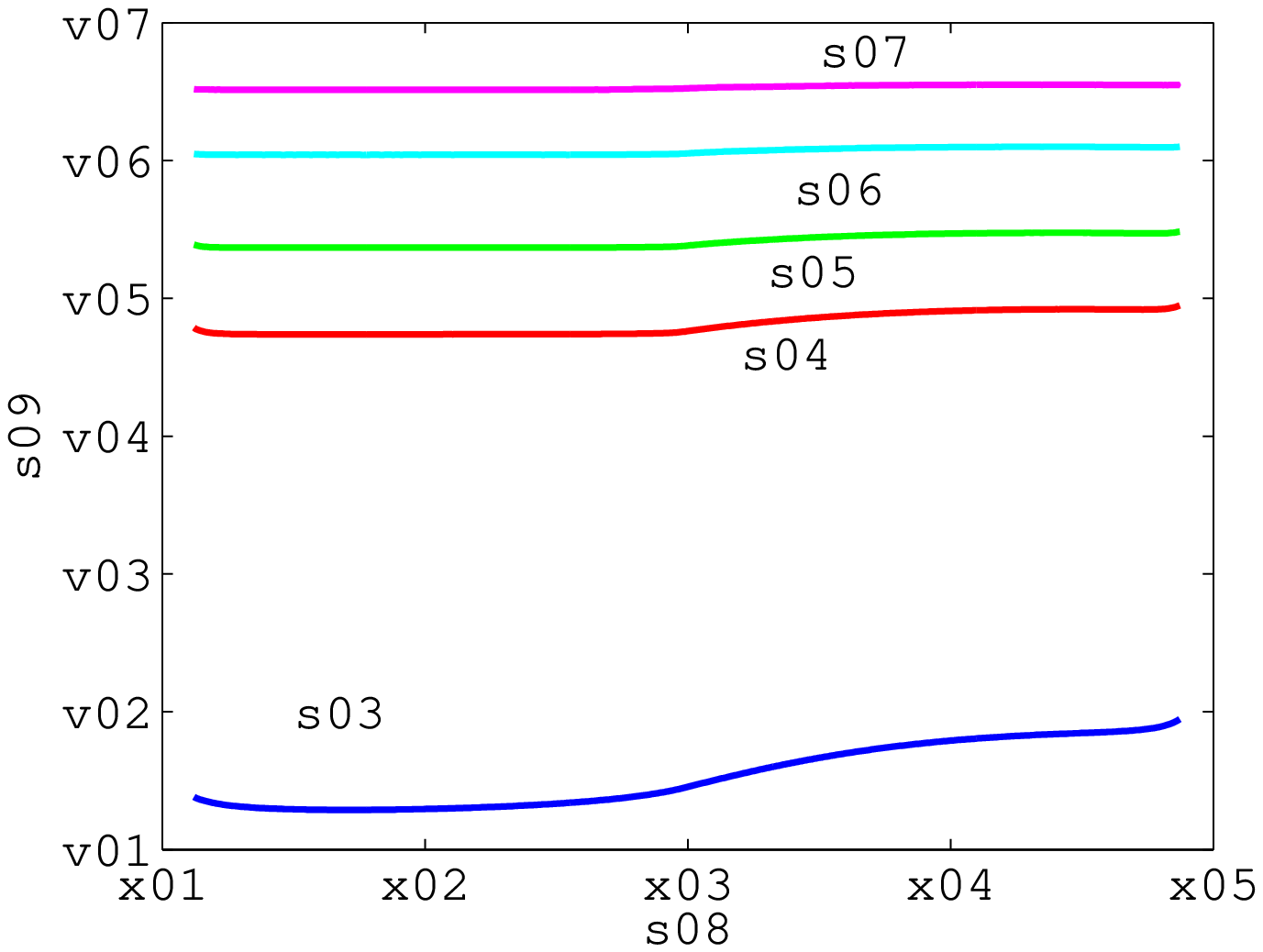}
\end{psfrags}%
%

%% file: fig9a.tex
%
%
\begin{psfrags}%
\psfragscanon%
%
\psfrag{s03}[l][l]{\color[rgb]{0,0,0}\setlength{\tabcolsep}{0pt}\begin{tabular}{l}a\end{tabular}}%
\psfrag{s04}[t][t]{\color[rgb]{0,0,0}\setlength{\tabcolsep}{0pt}\begin{tabular}{c}$\displaystyle \frac{z}{h/2}$ \end{tabular}}%
\psfrag{s05}[B][b]{\color[rgb]{0,0,0}\setlength{\tabcolsep}{0pt}\begin{tabular}{c}$(c_+ - c_{+,\infty})/c_{+,\infty}$\end{tabular}}%
%
\psfrag{x01}[t][t]{-1}%
\psfrag{x02}[t][t]{-0.8}%
\psfrag{x03}[t][t]{-0.6}%
\psfrag{x04}[t][t]{-0.4}%
\psfrag{x05}[t][t]{-0.2}%
\psfrag{x06}[t][t]{0}%
%
\psfrag{v01}[r][r]{-2}%
\psfrag{v02}[r][r]{0}%
\psfrag{v03}[r][r]{2}%
\psfrag{v04}[r][r]{4}%
\psfrag{v05}[r][r]{6}%
\psfrag{v06}[r][r]{8}%
%
\includegraphics[clip,width=3in]{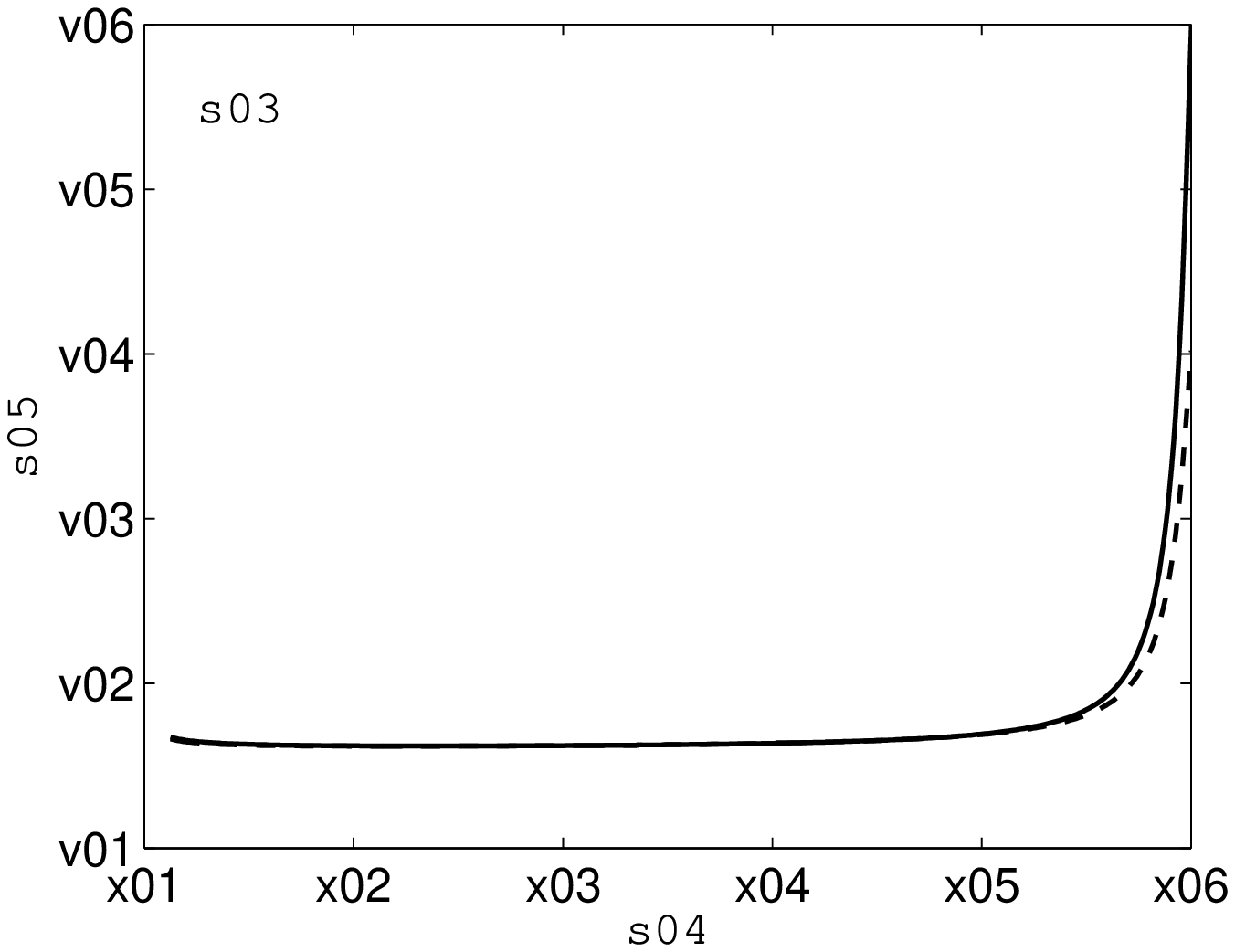}
\end{psfrags}%
%

%% file: fig9b.tex
%
%
\begin{psfrags}%
\psfragscanon%
%
\psfrag{s03}[l][l]{\color[rgb]{0,0,0}\setlength{\tabcolsep}{0pt}\begin{tabular}{l}b\end{tabular}}%
\psfrag{s04}[t][t]{\color[rgb]{0,0,0}\setlength{\tabcolsep}{0pt}\begin{tabular}{c}$\displaystyle \frac{z}{h/2}$ \end{tabular}}%
\psfrag{s05}[b][b]{\color[rgb]{0,0,0}\setlength{\tabcolsep}{0pt}\begin{tabular}{c}$c_{hi}^2/c_{low}^2$\end{tabular}}%
%
\psfrag{x01}[t][t]{-1}%
\psfrag{x02}[t][t]{-0.8}%
\psfrag{x03}[t][t]{-0.6}%
\psfrag{x04}[t][t]{-0.4}%
\psfrag{x05}[t][t]{-0.2}%
\psfrag{x06}[t][t]{0}%
%
\psfrag{v01}[r][r]{0.4}%
\psfrag{v02}[r][r]{}%
\psfrag{v03}[r][r]{0.6}%
\psfrag{v04}[r][r]{}%
\psfrag{v05}[r][r]{0.8}%
\psfrag{v06}[r][r]{}%
\psfrag{v07}[r][r]{1}%
\psfrag{v08}[r][r]{}%
%
\includegraphics[clip,width=3in]{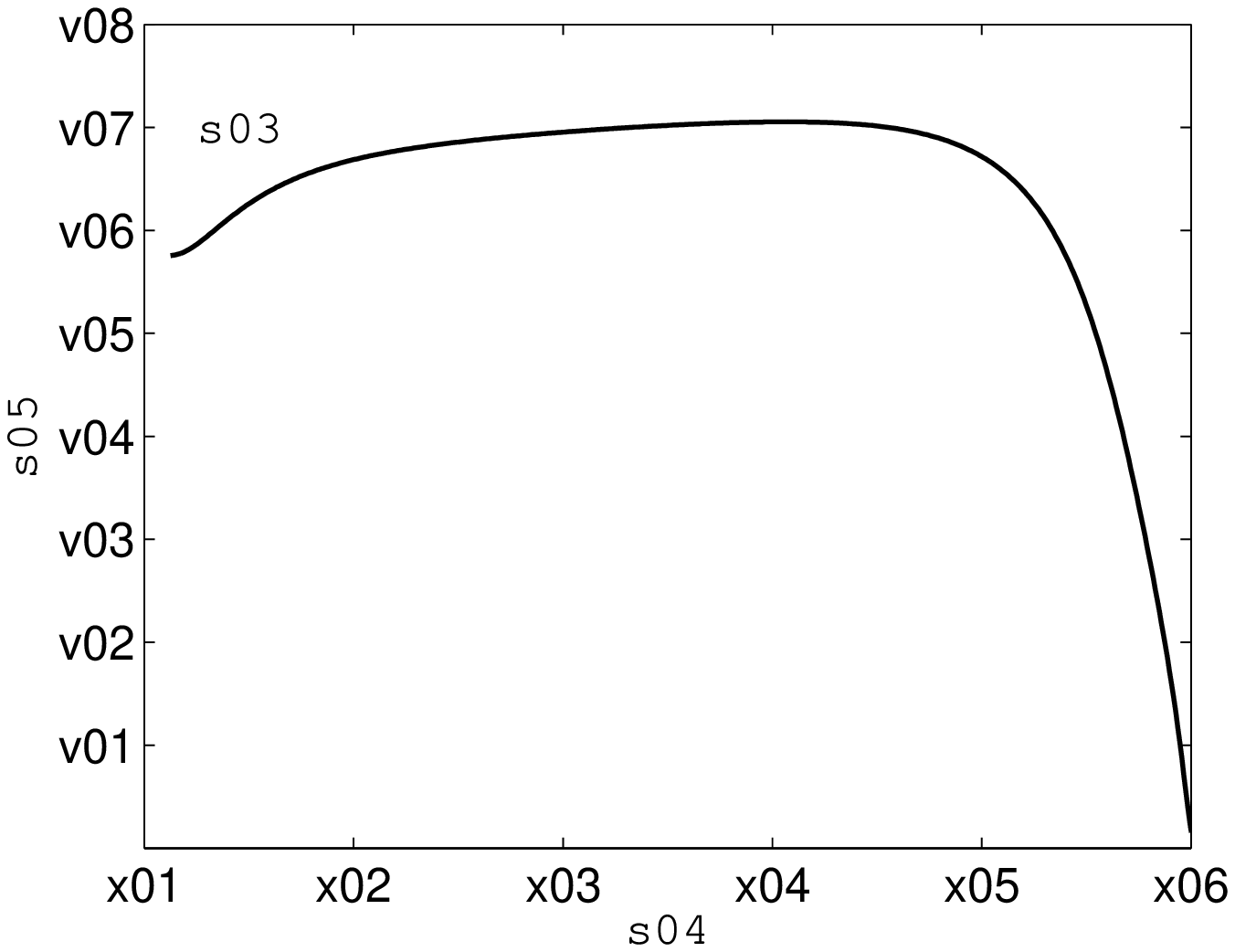}
\end{psfrags}%
%

%% file: salttab2.tex
\begin{table}

\begin{minipage}[c]{13cm}
\centering
\begin{tabular}{l c c}
\textbf{} \hspace{0.4cm} & \hspace{0.2cm} \textbf{Formula} \hspace{0.4cm} & \hspace{0.4cm} \textbf{Numerical Estimate} \\
\hline \hline
$Ra_e$ & $9.5 \times 10^{-5}$ / $9.2 \times 10^{-6}$ & 0.78 / 0.17 \\
$\beta_+$ & $6.5 \times 10^{-4}$ / $8.6 \times 10^{-5}$ & $7.3 \times 10^{-3}$ / $2.4 \times 10^{-3}$ \\
$Da_{anode}$ & 1.1 / 0.53 & 0.59 / 0.56 \\
$Da_{cathode}$ & 4.7 / 2.4 & 0.59 / 0.56 \\
\hline
\end{tabular}
\end{minipage}
\caption{Calculated and estimated values of the Rayleigh and Damk\"{o}hler numbers, as well as the parameter $\beta$ for protons, for two different conductivities.  The first column shows the calculated values of the dimensionless parameters from (\ref{Raedef}), (\ref{betadef}), (\ref{Da_anode_def}), and (\ref{Da_cathode_def}) for bulk conductivities of 8.8 and 35 $\mu$S/cm (corresponding to bulk KCl concentrations of 56.4 and 231 $\mu$mol/L, respectively).  The second column shows numerical estimates of the dimensionless parameters for both values of the conductivity.}
\label{salttab2}
\end{table}

%% file: summary.tex

The speed of bimetallic particles is reduced in high-conductivity solutions for several reasons.  The electric field in the diffuse layer near the rod decreases by approximately 90 \%, the total reaction rate by roughly 20 \%, and the rod potential by roughly 45 \%.  The swimming speed scales linearly with these three variables.

This work confirms the inverse relationship between the speed and ionic strength predicted by the scaling analysis, Eq. (\ref{scaling}).  However, the dependence of speed on ionic strength was measured not to be $U \propto C^{-1}$, but $U \propto C^{-1.3}$.  This deviation is a consequence of the fact that the reaction rate and zeta potential magnitude both decrease with ionic strength, further reducing the swimming speed.

Since we solve the full governing equations for the transport of charged species, our work automatically accounts for the presence of surface conduction in the EDL near the rod.  In externally driven electrophoresis, the role of surface conduction is most often characterized by the Dukhin-Bikerman number, $Du$, which is a measure of the surface conduction relative to the bulk.\cite{bikerman_structure_1942,dukhin_electrokinetic_1974,lyklema1995fundamentals,obrien_electrophoretic_1978,khair_influence_2009} In systems when the electric field is externally applied, the electrophoretic mobility decreases with increasing $Du$ (decreasing ionic strength).  In this paper, we show (see Figure S2) that the swimming speed increases with $Du$ (decreasing ionic strength).  The primary reason for this apparent contrast to externally driven electrophoresis is that the electric field in RICA is driven by surface reactions.  External electric fields are not typically current limited and if the ionic strength increases for a given electric field the total current also increases.   In RICA, the system has a finite current density $i$ that is roughly limited by the rate constants and reactant concentrations (see Eq. (\ref{fluxbc})) such that the electric field it generates $i/\sigma$ is  an  inverse function of the solution conductivity (or ionic strength).  In this way, as we add more salt the $Du$ decreases and so does the swimming velocity in contrast to electrophoresis.  We suggest that the Dukhin-Bikerman number does not aid in the understanding of RICA because of the proximity of the electric field that is generated by the surface reactions.


Although the simulations in this work account for the presence of the adsorbed Stern layer, the model of this layer can be made more detailed.  In particular, the zeta potential of the particle is likely influenced not only by the potential boundary condition, but also by the surface charge density associated with charged adsorbed species.  Future work could improve upon our model by including an adsorption isotherm in the model of the surface.  This adsorbed charge would affect the zeta potential, introducing an additional component to the potential boundary condition, Eq. (\ref{potentialBC}).

For simplicity, we have taken the permittivity here to be constant throughout the domain.  Additionally, the dielectric constant in the presence of the rod surface, where the electric fields are extremely strong, can decrease by as much as an order of magnitude due to the polarization of solvent molecules.\cite{delahay_double_1965,bard_electrochemical_2000,newman_electrochemical_2004}

Another aspect of the model with potential for improvement is the modeling of the reaction kinetics.  Although our previous work was the first to accurately describe the relationship between the reaction kinetics and rod potential,\cite{moran_electrokinetic_2011} it is generally accepted that the oxidation and reduction of hydrogen peroxide on platinum both occur through a series of intermediate steps.  However, Sabass and Seifert recently considered possible reaction pathways for peroxide oxidation and reduction, finding that both reactions were likely first-order in peroxide concentration, as was assumed here.\cite{sabass_nonlinear_2012}  Again for simplicity, here we have considered the peroxide oxidation and reduction reactions to be single-step processes on the Pt and Au surfaces, respectively.  Here our primary focus is to provide the first physically accurate description of the interfacial phenomena that lead to rod propulsion.

The results in this work suggest that the self-electrophoresis mechanism is inherently susceptible to speed reductions brought about by adding salt and are not practical for salt-rich environments.  Recently, several designs have been advanced for nanomotors which are capable of motion in salt-rich environments, using alternative fuels, and novel high-speed designs.  From the results of this study, these latter particles seem the most likely candidates to realize the autonomous swimming in electrolytes such as biological media.\cite{gao_highly_2011,manesh_template_2010,gao_water-driven_2012,gao_hydrogen-bubble-propelled_2012} 